\documentclass[11pt]{article}
\usepackage{amssymb,amsmath,amsthm,mathtools}
\usepackage{array,geometry}  
\usepackage{setspace}
\usepackage{bm,paralist,dsfont,nicefrac}
\usepackage{etoolbox}
\usepackage[normalem]{ulem}
\usepackage{palatino}
\usepackage[T1]{fontenc}

\usepackage{tikz,float,booktabs,pifont,multirow,subcaption}
\usepackage{makecell}
\usetikzlibrary{fit,calc,math,shapes,shapes.multipart,decorations.text,arrows,decorations.markings,decorations.pathmorphing,shapes.geometric,positioning,decorations.pathreplacing, patterns}

\usepackage{natbib}

\usepackage[pagebackref,colorlinks]{hyperref}
\hypersetup{linkcolor=[rgb]{.7,0,0}}
\hypersetup{citecolor=[rgb]{0,.5,0}}
\hypersetup{urlcolor=[rgb]{.7,0,.7}}
\renewcommand*{\backref}[1]{}
\renewcommand*{\backrefalt}[4]{%
    \ifcase #1 (Not cited.)%
    \or        (Cited on page~#2)%
    \else      (Cited on pages~#2)%
    \fi}

\usepackage[capitalize, nameinlink, noabbrev]{cleveref} 

\AddToHook{cmd/appendix/before}{%
	\crefalias{section}{appendix}%
	\crefalias{subsection}{appendix}
}

\usepackage{xcolor}

\usepackage{thmtools,thm-restate}
\usepackage{enumitem}
\usepackage{nicematrix}

\usepackage{longtable}
\usepackage{booktabs}

\allowdisplaybreaks

\newcommand{\EFX}{\textrm{\textup{EFX}}}
\newcommand{\EFone}{\textrm{\textup{EF1}}}

\renewcommand{\epsilon}{\varepsilon}

\newcommand{\MMS}[1]{\ifstrempty{#1}{\textrm{\textup{MMS}}}{\textrm{${#1}$-\textup{MMS}}}}

\newcommand{\V}{\mathcal{V}}

\newcommand{\MatchAndPause}{\textup{\textsc{MatchAndPause}}}

\newcommand{\PHASE}[1]{\item[]{\color{gray}\textit{$\triangleright$ #1}}}

\usepackage{algorithm}
\usepackage{algorithmic}

\newtheorem{theorem}{Theorem}
\newtheorem*{theorem*}{Theorem}

\newtheorem*{definition*}{Definition}

\newtheorem{lemma}{Lemma}
\newtheorem*{lemma*}{Lemma}

\newtheorem*{claim*}{Claim}

\newtheorem*{fact*}{Fact}

\newtheorem*{observation*}{Observation}

\newtheorem*{conjecture*}{Conjecture}

\newtheorem*{corollary*}{Corollary}

\newtheorem*{remark*}{Remark}

\newtheorem{proposition}{Proposition}
\newtheorem*{proposition*}{Proposition}

\newtheorem*{example*}{Example}

\Crefname{algocf}{Algorithm}{Algorithms}

\usepackage{colortbl}
\colorlet{myred}{red!25}
\colorlet{myblue}{blue!25}
\colorlet{mygreen}{green!25}

\usepackage{todonotes}

\usepackage{tikz}
\usetikzlibrary{arrows.meta,decorations.pathreplacing}

\definecolor{largegood}{HTML}{D5E6F0}
\definecolor{smallgood}{HTML}{F8D7A8}
\definecolor{ink}{HTML}{253747}
\definecolor{reserved}{HTML}{795697}

\title{The Golden Ratio: \EFone{} and Ordinal-\MMS{} for Additive Goods }

\author{
 \begin{tabular}{m{0.12\linewidth}m{0.12\linewidth}m{0.12\linewidth}m{0.12\linewidth}m{0.12\linewidth}m{0.12\linewidth}}
 \multicolumn{3}{c}{\textbf{Hadi Hosseini}} & \multicolumn{3}{c}{\textbf{Shraddha Pathak}} \\
          \multicolumn{3}{c}{\small{Pennsylvania State University}} & \multicolumn{3}{c}{\small{Pennsylvania State University}} \\
          \multicolumn{3}{c}{\href{mailto:hadi@psu.edu}{\small{\texttt{hadi@psu.edu}}}} & \multicolumn{3}{c}{\href{mailto:ssp5547@psu.edu}{\small{\texttt{ssp5547@psu.edu}}}} \\
   \end{tabular}
}

\date{}

\begin{document}

\maketitle 

%%%%%%%%%%%%%%%%%%%%%%%%%%%%%%%%%%%%%%%%%%%%%%%%%%%%%%%%%%%%%%%%%%%%%%%%

\begin{abstract}
    Every complete allocation that is envy-free up to one good (\EFone{}) guarantees each of the $n$ agents her $1$-out-of-$(2n-1)$ maximin share. Can we construct allocations that satisfy \EFone{} together with stronger ordinal maximin share guarantees? For additive valuations, we prove that there always exists a complete allocation that is simultaneously \EFone{} and  ($1$-out-of-$\lfloor\varphi n\rfloor$)-\MMS{}, where $\varphi=(1+\sqrt{5})/2$ is the golden ratio.
    For bivalued valuations, where each agent may have her own two non-negative values for the goods, we obtain a complete allocation that is simultaneously envy-free up to any good (\EFX{}) and ($1$-out-of-$\lceil(7n-1)/6\rceil$)-\MMS{}.
\end{abstract}

\section{Introduction}
\label{sec:Introduction}

The fair division literature studies a wide range of often orthogonal fairness notions. 
Pairwise guarantees compare an agent's bundle with those assigned to other agents, while threshold-based guarantees require her to receive a certain share of the available value. These notions capture different aspects of fairness and need not imply one another: an allocation that satisfies one may fail another. This raises a compatibility question: can pairwise and threshold-based fairness guarantees be satisfied simultaneously?

Two well-known notions capturing these perspectives are envy-freeness up to one good (\EFone{}) and maximin share (\MMS{}), respectively. An allocation is \EFone{} if removing some good from an envied bundle eliminates envy \citep{LMM+04approximately,B11combinatorial}. 
\MMS{} requires every agent to receive a bundle of value at least the value she can guarantee by partitioning the goods into $n$ bundles and receiving the least-valuable bundle, where $n$ is the number of agents \citep{B11combinatorial}.
While an \EFone{} allocation is always guaranteed to exist, solutions giving every agent her maximin share may not exist---even with additive valuations and only three agents \citep{KPW18fair}.
The non-existence of \MMS{}, together with its computational intractability \citep{KPW18fair} have motivated the design of approximation guarantees of \MMS{}.

We focus on simultaneously satisfying \EFone{} and achieving ordinal approximations of \MMS{}. An agent's $1$-out-of-$d$ maximin share is the value she can guarantee for herself by partitioning the goods into $d\ge n$ bundles and receiving the least-valued bundle in the partition.

There are several technical challenges in satisfying \EFone{} and approximations of \MMS{}.
A common algorithmic technique for approximating \MMS{} utilizes ordering-unordering of an instance, while preserving the \MMS{} threshold, for example \citep{BK20approximation,AG24breaking,HSS22ordinal,AGST24improving}.
While effective, such transformations do not generally preserve \EFone{}.
Other direct approaches that do not rely on ordering instances, e.g. \citet{KPW18fair}, have been utilized to achieve multiplicative approximation of $\nicefrac{2}{3}$-\MMS{}, and later adapted to give simultaneous \EFone{} and $\nicefrac{2}{3}$-\MMS{} guarantees. 

For ordinal approximations, every complete \EFone{} allocation directly give $1$-out-of-$(2n-1)$ maximin share guarantee.
In this paper, we ask: \emph{Can we construct a complete allocation that simultaneously satisfies \EFone{} and ($1$-out-of-$d$)-\MMS{} for some $d<2n-1$?}

\subsection{Contributions}

\paragraph{General Additive Valuations.}

For general additive valuations, we show that there always exists a complete allocation that simultaneously satisfies \EFone{} and $1$-out-of-$\lfloor\varphi n\rfloor$-\MMS{}, where $\varphi=(1+\sqrt5)/2$ is the golden ratio (\Cref{thm:golden-ratio-ef1}). For $n\ge 3$, this strengthens the $1$-out-of-$(2n-1)$ guarantee that follows from \EFone{} alone. Our construction uses a divider procedure (\textsc{Shrink-and-Match}; \Cref{alg:divider}) in which one agent forms bundles that meet her own demand, and a matching assigns some of them to the agents who find them acceptable. The procedure maintains \EFX{} among agents whose demands have been met and ensures that each remaining agent values every assigned bundle below her demand.

The challenge is to ensure that there is some divider who can always form enough such bundles from the remaining goods. For agents whose ordinal targets do not already follow from \EFone{}, we try to reserve distinct valuable goods to reduce their demands. If this is not possible, Hall's theorem identifies a group of agents who can serve as dividers until enough bundles have been assigned for a greedy procedure to suffice. 
Our construction works directly with the original valuations. This is important since the standard reduction to an ordered instance, in which all agents agree on the ranking of goods, preserves ordinal \MMS{} guarantees but need not preserve \EFone{} when transferring the allocation back to the original instance.

\paragraph{Bivalued Valuations.}
Under bivalued valuations, where each agent $i\in N$ values every good at either $a_i$ or $b_i$, we devise an algorithm that achieves stronger guarantees of \EFX{} and $1$-out-of-$\lceil(7n-1)/6\rceil$-\MMS{} (\Cref{thm:bv-main}).
In each round, \MatchAndPause{} matches as many active agents as possible to goods they value highly. It then temporarily pauses certain matched agents, giving those who missed out on a high-valued good a chance to catch up by receiving low-valued goods.
The proof considers what happens in the last round in which a good that a given agent values highly is allocated, requiring a detailed analysis based on the number of high and low goods in each bundle as well as the use of auxiliary weights.
The algorithm runs in polynomial time when all values are rational.

\subsection{Related Work}

\paragraph{Envy guarantees.}
\EFone{} allocations always exist for additive valuations and can be computed in polynomial time (even for all monotone valuations) using the envy-cycle elimination algorithm of \citet{LMM+04approximately}. It can also be simultaneously satisfied with certain efficiency notions e.g. \citet{CKM+19unreasonable} showed that maximum Nash welfare allocations satisfy both \EFone{} and Pareto optimality. The existence of complete \EFX{} allocations remains open for general additive valuations. Positive results include instances with restrictions on either the number of agents or the type of valuations. \citet{PR20almost} show the existence of \EFX{} for two agents and \citet{CGM24efx} extend it to three agents. \EFX{} allocations also exist for instances with arbitrary number of agents with identical valuations~\citep{PR20almost} or bivalued valuations~\citep{ABFHV21maximum}. \citet{BMP25probing} and \citet{JT25personalized} gave polynomial-time algorithms for \EFX{} under personalized bivalued valuations where the two values may differ across agents. In particular, \citet{BMP25probing} adapted \citet{ABFHV21maximum}'s Match-and-Freeze algorithm to this setting. Our bivalued algorithm follows this approach, and our analysis establishes an ordinal \MMS{} guarantee for the allocation it produces.

\paragraph{Maximin share guarantees.}
Allocations where every agent receives a bundle of value at least her maximin share need not exist, even with additive valuations and only three agents~\citep{KPW18fair}. For personalized bivalued valuations, exact \MMS{} allocations can be computed in polynomial time when each agent's high value is an integer multiple of her positive low value~\citep{EPS22easy}, and a recent result extends this to arbitrary ratios~\citep{Z26exact}. However, the non-existence for general additive has motivated the study of two types of relaxations of \MMS{}: multiplicative approximations and ordinal approximations. Multiplicative approximations guarantee each agent a fixed fraction of her maximin share. \citet{KPW18fair} established a $2/3$ guarantee, and subsequent work improved the factor beyond $3/4$ to $3/4+3/3836$~\citep{AG24breaking}. More recently, \citet{HZ26strongly} obtained a $7/9$ guarantee.

Ordinal approximations of \MMS{} use partitions into $d\ge n$ bundles to calculate the threshold values (the allocation is still among $n$ agents). \citet{AHS22envy} established the existence of $1$-out-of-$(2n-2)$-\MMS{} allocations. Polynomial-time algorithms for small $n$ were first given by \citet{hosseini2021guaranteeing}. For arbitrary $n$, \citet{HSS22ordinal} improved the guarantee to ($1$-out-of-$\lfloor 3n/2\rfloor$)-\MMS{} and gave a polynomial-time algorithm for ($1$-out-of-$\lceil 3n/2\rceil$)-\MMS{}.  \citet{AGST24improving} subsequently proved the existence of $1$-out-of-$4\lceil n/3\rceil$-\MMS{} allocations. An advantage of ordinal guarantees is that they depend only on how an agent ranks bundles. Changing numerical values while preserving these rankings preserves whether a bundle meets the guarantee, whereas a multiplicative guarantee can be sensitive to such changes~\citep{HSS22ordinal}. Depending on the instance, either type of relaxation can give the higher threshold.

\paragraph{Simultaneous fairness guarantees.}
Several recent works study allocations that satisfy multiple requirements at the same time; some that consider fairness with efficiency \citep{CKM+19unreasonable,BKV18Finding} and some that consider different fairness notions themselves \citep{AMN20multiple,AR25achieving,AR26simultaneous,hosseini2026simultaneous}. 
Under natural restrictions, \EFone{} can be satisfied together with exact \MMS{}; example, for identical valuations \citep{PR20almost}, binary additive valuations \citep{SS19democratic}, and weakly lexicographic preferences \citep{hosseini2024almost}.
Recently, \citet{AR25achieving} showed that every general additive instance admits an allocation satisfying both \EFone{} and $2/3$-\MMS{}. Some of our construction in \Cref{sec:general-additive} (\textsc{Shrink-and-Match}; \Cref{alg:divider}) builds on their approach, specifically the one related to obtaining envy-based guarantees; the challenge for us is to obtain ordinal-\MMS{} guarantees within their framework.

Other work uses the residual maximin share to obtain partial allocations satisfying \EFX{} and complete allocations satisfying envy-freeness up to one less-preferred good (EFL)~\citep{F25residual,akrami2026power}.
\citet{AAM26polynomial} give polynomial-time algorithms combining EFL with approximations of the AnyPrice share and the minimum \EFX{} share.
The closest results to ours are by \citet{AR26simultaneous} who study simultaneous ordinal and envy guarantees. They show that for ordered instances, in which all agents rank the goods in the same order, \EFX{}+($1$-out-of-$\lceil 3n/2\rceil$)-\MMS{}, and \EFone{}+($1$-out-of-$4\lceil n/3\rceil$)-\MMS{} can be obtained. When agents agree only on the set of their $n$ most valuable goods, they obtain \EFone{} together with $1$-out-of-$\lceil 3n/2\rceil$-\MMS{}.
As opposed to these restrictions, in this work, we focus on the general additive setting. 

\section{Preliminaries}
\label{sec:Preliminaries}

For an integer $k\ge 1$, we write $[k]\coloneqq \{1, 2, \ldots, k\}$.

\paragraph{Problem instance.}
An \emph{instance} consists of a finite set $N$ of agents (with $n=|N|\ge 1$), a finite set $M$ of indivisible goods (where $m=|M|$), and a valuation profile $\V=(v_1,\ldots,v_n)$, where $v_i:2^M\to\mathbb{R}_{\ge 0}$ and $v_i(\varnothing)=0$. 
Throughout, we assume that every valuation function $v_i$ is \emph{additive} i.e. $v_i(S)=\sum_{g\in S}v_i(g)$ for every $S\subseteq M$, where, for simplicity, $v_i(g) \coloneqq v_i(\{g\})$. 

An instance is said to have \emph{bivalued} valuations if for every $i$, there exists $a_i, b_i \in \mathbb{R}_{\ge 0}$ such that agent $i$ values every good $g\in M$ at either of these two values i.e. $v_i(g)\in \{a_i, b_i\}$ for every $g\in M$.

An \emph{allocation} is a tuple $A=(A_1,\ldots,A_n)$ of pairwise disjoint subsets of $M$, where agent $i$ receives the bundle $A_i$.
We say that an allocation is \emph{partial} if $\cup_{i\in N}A_i \subsetneq M$ and it is \emph{complete} if $\cup_{i\in N} A_i = M$.

\paragraph{Envy-based fairness notions.}
An allocation $A$ is said to be envy-free if, for every pair of agents $i,j\in N$, we have $v_i(A_j)\le v_i(A_i)$. Allocation $A$ is envy-free up to one good (\EFone{}) if for every pair of agents $i,j\in N$ where $v_i(A_i)< v_i(A_j)$, there exists a good $g\in A_j$ such that $v_i(A_i) \ge v_i(A_j\setminus \{g\})$. Allocation $A$ is envy-free up to any good (\EFX{}) if for every pair of agents $i,j\in N$ and every good $g\in A_j$, we have $v_i(A_i) \ge v_i(A_j\setminus \{g\})$\footnote{This definition allows for the removal of zero-valued goods, and is sometimes referred to as \EFX{}$_0$.}.
When there exists some $g\in A_j$ such that $v_i(A_i) < v_i(A_j\setminus \{g\})$, we say that $i$ \EFX{}-envies or strongly envies the bundle $A_j$.

\paragraph{Share-based fairness notions.}
Let $\Pi_d(M)$ denote the set of partitions of $M$ into $d$ bundles, allowing empty bundles. 
For $d\ge n$, agent $i$'s ($1$-out-of-$d$)-maximin share is defined as
\[\mu_i^d \coloneqq \max_{(P_1, \ldots, P_d) \in \Pi_d(M)} \min_{r \in [d]} v_i(P_r).\]
An allocation $A$ is said to be ($1$-out-of-$d$)-\emph{\MMS{}} if $v_i(A_i)\ge\mu^d_i$ for every $i\in N$.
When $d= n$, we simply say that an allocation $A$ is \MMS{} and refer to its corresponding share as $\mu_i$. For $d>n$ , the notion is weaker than \MMS{} since $\mu_i^d\le \mu_i$.

\paragraph{Graphs and matchings.}
For a bipartite graph $G=(X,Y;E)$ and a set $Z$ of vertices on one side, let $\Gamma_G(Z)$ denote the vertices on the other side adjacent to at least one vertex in $Z$.
We omit the subscript when the graph is clear from context.

A \emph{matching} is a set of edges with no shared endpoints. A vertex is \emph{matched} if it is incident to a matching edge, and is \emph{unmatched} otherwise. A matching \emph{covers} a set of vertices if every vertex in that set is matched; it is \emph{perfect} if it covers all vertices.
A maximum-cardinality matching has the largest possible number of edges. By Hall's theorem, $G$ has a matching covering $X$ if and only if $|\Gamma_G(Z)|\ge |Z|$ for every $Z\subseteq X$.

Given a matching $Q$, an \emph{alternating path} is a path whose edges alternate between edges outside $Q$ and edges in $Q$. An \emph{augmenting path} is an alternating path with two unmatched endpoints. Exchanging the matching and non-matching edges along an augmenting path increases the size of the matching by one. Thus, a maximum-cardinality matching has no augmenting path.

\section{Additive Valuations}
\label{sec:general-additive}

In this section, we show that an allocation that is simultaneously \EFone{} and ($1$-out-of-$\lfloor \varphi n\rfloor$)-\MMS{} always exists for additive valuations, where $\varphi=(1+\sqrt5)/2$ is the golden ratio.

\paragraph{Overview of the construction.}

Each agent has a \emph{demand}: either her ordinal threshold or the value she still needs after a good has been reserved for her.
We define a \textsc{Shrink-and-Match} algorithm where, in every round, a divider forms as many disjoint bundles as there are remaining agents, each of value at least her demand.
We shrink bundles, allow satisfied agents to exchange their bundles for subsets of these proposed bundles. Together with a suitable matching, these steps ensure that (a) each agent left unsatisfied values every assigned bundle below her demand, and (b) all satisfied agents meet their demand and are \EFX{} towards one another. This part of the construction builds on \citet{AR25achieving}.
It remains to show that a suitable divider can always
form enough bundles. 

We identify agents whose targets follow from any complete \EFone{} allocation. For the others, we try to reserve distinct valuable goods, called \emph{anchors}. If suitable anchors can be assigned, we use \textsc{Shrink-and-Match} (\Cref{alg:divider}) to meet the remaining demands. The anchors are chosen so that appending them gives \EFone{}. Otherwise, Hall's theorem identifies a group of agents with few valuable goods who can divide during the early iterations. Once sufficiently many agents are satisfied, greedy packing allows any remaining agent to divide.

\begin{figure}[t]
\centering
\begin{tikzpicture}[
    font=\small,
    box/.style={
        draw=black!65,
        rounded corners=2pt,
        align=center,
        text width=2.05cm,
        minimum height=1.35cm,
        inner sep=5pt,
        line width=0.6pt
    },
    branch/.style={
        box,
        text width=2.35cm,
        minimum height=1.95cm
    },
    line/.style={
        draw=black!65,
        line width=0.6pt
    },
    arrow/.style={
        line,
        -{Stealth[length=1.8mm]}
    }
]

% Decision
\node[box] (decision) at (0,0)
    {Can suitable anchors be reserved?};

% Two ways to choose demands and dividers
\node[branch, fill=black!4] (anchors) at (3.1,1.2)
    {Reserve anchors\\[3pt]
     Use remaining demands};

\node[branch, fill=black!4] (hall) at (3.1,-1.2)
    {Original targets\\[3pt]
     Hall's theorem};

% Shared subroutine
\node[box, text width=3.5cm]
    (subroutine) at (7,0)
    {Run\\[3pt]
     \textsc{Shrink-and-Match}};

% Final steps
\node[box, text width=1.8cm] (append) at (10.4,0)
    {Append anchors, if reserved};

\node[box, text width=2.1cm] (complete) at (13.2,0)
    {Complete by envy-cycle elimination};

% Split
\coordinate (split) at (1.5,0);
\draw[line] (decision.east) -- (split);
\draw[arrow] (split) |- (anchors.west);
\draw[arrow] (split) |- (hall.west);

\node[font=\scriptsize, above] at (1.5,1.2) {Yes};
\node[font=\scriptsize, below] at (1.5,-1.2) {No};

% Merge into the shared subroutine
\coordinate (merge) at (4.8,0);
\draw[line] (anchors.east) -| (merge);
\draw[line] (hall.east) -| (merge);
\draw[arrow] (merge) -- (subroutine.west);

\draw[arrow] (subroutine.east) -- (append.west);
\draw[arrow] (append.east) -- (complete.west);

\end{tikzpicture}
\caption{Overview of the construction.
We seek suitable anchors for agents whose ordinal targets are not already guaranteed by a complete \EFone{} allocation. If anchors can be reserved, \textsc{Shrink-and-Match} (\Cref{alg:divider}) meets the remaining demands using non-anchor goods. Otherwise, demands are the original ordinal targets: Hall's theorem identifies suitable early dividers, and greedy packing allows any remaining agent to divide once sufficiently many agents are satisfied.
After appending any reserved anchors, envy-cycle elimination completes the allocation while preserving \EFone{} and without decreasing any agent's value.}
\label{fig:additive-overview}
\end{figure}
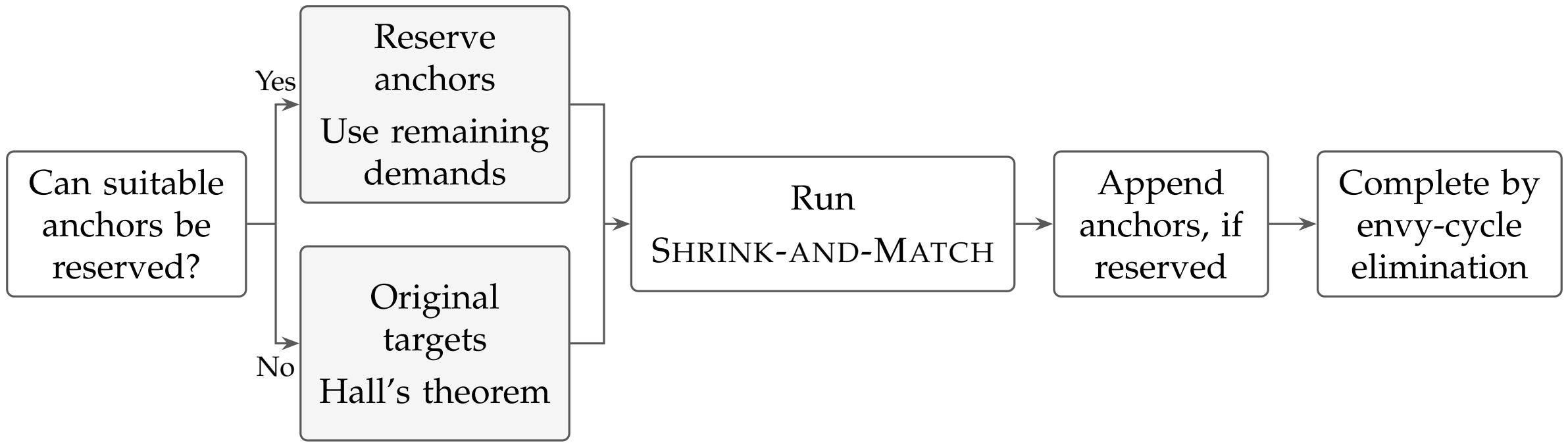

In both cases, envy-cycle elimination completes the allocation while preserving \EFone{} and without decreasing any agent's value \citep{LMM+04approximately}. \Cref{fig:additive-overview} illustrates the construction.

\subsection{\textsc{Shrink-and-Match}: An \EFX{} Divider Procedure}
\label{subsec:divider}

The procedure takes as input a set of goods $U_0\subseteq M$ and a demand $\tau_i\ge 0$ for each agent $i\in N$. Initially, all goods in $U_0$ are available. Throughout the procedure, $U$ denotes the set of goods from $U_0$ that are currently unassigned. A bundle $B$ is \emph{acceptable} to agent $i$ if $v_i(B)\ge\tau_i$. In one case, the demand is directly the agent's ordinal target. In another, it is the additional value she needs after a good has been reserved for her.

At any intermediate step, $S\subseteq N$ denotes the set of satisfied agents whose demands have been met, and let the remaining agents be $N'=N\setminus S$. Agents in $N'$ hold empty bundles.
Throughout the execution of the algorithm, we have
\begin{equation}
    v_i(A_j)<\tau_i
    \qquad\text{for every }i\in N'\text{ and }j\in S.
    \label{eq:unassigned-demand}
\end{equation}
That is, each remaining agent considers every assigned bundle insufficient to meet her own demand.

\paragraph{Divide and match.}
Suppose that a divider from $N'$ can form $|N'|$ pairwise disjoint bundles $\mathcal{B}$ from the unassigned goods $U$, each acceptable to her.
We construct a bipartite graph with vertex sets $N'$ and $\mathcal{B}$, where there is an edge between agent $i$ and bundle $B$ if $v_i(B)\ge\tau_i$.
We want to find a matching in this graph where no unmatched agent is adjacent to a matched bundle.
Assigning bundles according to such a matching satisfies its recipients and preserves~\eqref{eq:unassigned-demand} for the agents who remain in $N'$.

\begin{lemma}
\label{lem:envy-free-matching}
Let $G=(N',\mathcal{B};F)$ be a bipartite graph with $|N'|=|\mathcal{B}|\ge 1$.
If every vertex in $\mathcal{B}$ has a neighbor, then $G$ admits a non-empty matching where no unmatched agent is adjacent to a matched bundle.
\end{lemma}

We defer the proof to \Cref{sec:appendix-general-additive}.
The lemma applies whenever the divider produces $|N'|$ bundles that are all acceptable to her.
It also continues to apply after shrinking the bundles, provided that each candidate remains acceptable to at least one agent in $N'$.

\paragraph{Minimal bundles.}
Before assigning a bundle, we shrink it to a subset $D$ that is acceptable to at least one agent in $N'$, but whose proper subsets are unacceptable to every agent in $N'$.
In particular, $v_i(D\setminus\{g\})<\tau_i$ for every $i\in N'\text{ and }g\in D.$
    
We obtain such a bundle by repeatedly removing a good whenever the remaining bundle is still acceptable to some agent in $N'$.
The removed goods remain unassigned.
Notice that minimality is with respect to all agents in $N'$, rather than only the divider or an intended recipient.

If an agent $i\in N'$ receives an acceptable bundle, her value is at least $\tau_i$.
By~\eqref{eq:unassigned-demand}, she does not envy any previously assigned bundle.
For every other candidate bundle $D$, minimality gives $v_i(D\setminus\{g\})<\tau_i$ for every $g\in D$.
Thus, she does not strongly envy any bundle assigned in the same round.

What remains is to control the envy from agents who are already satisfied. 
In particular, in addition to~\eqref{eq:unassigned-demand}, we maintain that the satisfied agents hold a bundle of value at least their demand and that they are \EFX{} towards every other satisfied agent i.e. 
\begin{align}
    v_i(A_i)&\ge\tau_i
    &&\text{for every }i\in S,
    \label{eq:satisfied-demand}\\
    v_i(A_i)&\ge v_i(A_j\setminus\{g\})
    &&\text{for every }i,j\in S
      \text{ and }g\in A_j.
    \label{eq:satisfied-envy}
\end{align}
Towards this, we combine minimal bundles with exchanges that improve the value of a satisfied agent.

\paragraph{Improving a satisfied agent.}
Suppose a satisfied agent $j\in S$ \EFX{}-envies a bundle $D$ i.e. for some $g\in D$, $v_j(D\setminus\{g\})>v_j(A_j).$
We choose an inclusion-wise minimal subset $B\subseteq D\setminus\{g\}$ that strictly improves at least one satisfied agent.
Thus, there exists a satisfied agent $h\in S$ with $v_h(B)>v_h(A_h)$, whereas for every $i\in S\text{ and }b\in B$, we have $v_i(B\setminus\{b\})\le v_i(A_i)$.
We then give $B$ to agent $h$, return her previous bundle to the unassigned goods, and restart the division step.
Since agent $h$ strictly improves, her demand remains satisfied and her envy comparisons with the unchanged bundles also remain valid.

\paragraph{The complete procedure.}
\Cref{alg:divider} combines these steps. The choice of divider and the bundles she proposes may vary. For now, we assume that she can form as many bundles that are acceptable to her as there are remaining agents.

\begin{algorithm}[t]
\caption{\textsc{Shrink-and-Match}}
\label{alg:divider}
\begin{algorithmic}[1]
\REQUIRE Agents $N$, available goods $U_0$, demands $(\tau_i)_{i\in N}$, and a division rule $\mathcal R$
\ENSURE A possibly partial \EFX{} allocation meeting every agent's demand

\STATE $U\gets U_0$, \quad $A_i\gets\varnothing$ for all $i\in N$, \quad $S\gets\{i\in N:\tau_i=0\}$
\WHILE{$S\ne N$}
    \STATE $N'\gets N\setminus S$
    \STATE Use $\mathcal R$ to select a divider $p\in N'$ and form $|N'|$ disjoint bundles from $U$, each acceptable to $p$
    \STATE Shrink each bundle to an inclusion-wise minimal subset acceptable to at least one agent in $N'$. Let $\mathcal B$ be the resulting collection.

    \IF{there exist $j\in S$, $D\in\mathcal B$, and $g\in D$ with $v_j(D\setminus\{g\})>v_j(A_j)$}
        \STATE Choose an inclusion-wise minimal $B\subseteq D\setminus\{g\}$ that strictly improves some agent in $S$
        \STATE Choose $h\in S$ with $v_h(B)>v_h(A_h)$
        \STATE Return $A_h$ to $U$, remove $B$ from $U$, and set $A_h\gets B$. Restart with a new division of $U$.
    \ELSE
        \STATE Form the bipartite graph on $N'$ and $\mathcal B$, joining each agent to the bundles she finds acceptable
        \STATE Choose a nonempty matching $P$ such that no unmatched agent is adjacent to a matched bundle (\Cref{lem:envy-free-matching})
        \FOR{each $(i,D)\in P$}
            \STATE $A_i\gets D$, \quad $U\gets U\setminus D$,
                   \quad $S\gets S\cup\{i\}$
        \ENDFOR
    \ENDIF
\ENDWHILE
\STATE \textbf{return} $(A_i)_{i\in N}$
\end{algorithmic}
\end{algorithm}

\begin{lemma}
\label{lem:divider-guarantee}
Suppose that, at every iteration, the selected divider can form $|N'|$ pairwise disjoint bundles from $U$, each acceptable to her.
Then the above procedure terminates after finitely many iterations and returns a (possibly partial) allocation $A$ such that, (i) for every $i\in N$, we have $v_i(A_i) \ge \tau_i$, and (ii) $A$ satisfies \EFX{}. 
\end{lemma}

\begin{proof}
We first show that \eqref{eq:unassigned-demand}, \eqref{eq:satisfied-demand}, and \eqref{eq:satisfied-envy} hold throughout the procedure.
Initially, every bundle is empty, agents in $S$ have zero demand, and agents in $N'$ have strictly positive demand; thus, all three conditions hold.

First, let us consider the case where an exchange assigns $B$ to an agent $h\in S$.
Agent $h$ strictly improves, and the bundles of all other agents remain unchanged.
Hence,~\eqref{eq:satisfied-demand} holds.
Since $B$ is a proper subset of a minimal acceptable bundle $D$, every agent $i\in N'$ has $v_i(B)<\tau_i$.
Thus,~\eqref{eq:unassigned-demand} also holds.
By the minimality of $B$, for every $i\in S$ and $b\in B$, the value $v_i(B\setminus\{b\})$ is at most agent $i$'s value before the exchange, so no satisfied agent \EFX{} envies the new bundle $B$.
Moreover, agent $h$ does not have \EFX{} envy towards any unchanged bundle because her own value increases. Therefore,~\eqref{eq:satisfied-envy} still holds.

Next, let us consider the case when bundles are assigned to a non-empty set of agents $H\subseteq N'$ in the matching step.
Each agent in $H$ receives an acceptable bundle, so \eqref{eq:satisfied-demand} holds for the larger set $S\cup H$.
The matching is such that there is no edge between an agent in $N'\setminus H$ and a newly assigned bundle.
Since the procedure reached the matching step, no previously satisfied agent strongly envies any newly assigned bundle.
For each newly satisfied agent $i\in H$, her own value is at least $\tau_i$.
She values every previously assigned bundle below $\tau_i$ and by minimality of the bundles,
\[
    v_i(D\setminus\{g\})<\tau_i
    \qquad
    \text{for every }D\in\mathcal{B}\text{ and }g\in D,
\]
she does not \EFX{} envy another newly assigned bundle.

To prove the lemma, it remains to show that the procedure terminates in finitely-many steps.
Every matching update increases $|S|$, so there are at most $n$ such updates.
Between two matching updates, the set $S$ is fixed.
Each exchange strictly increases one agent's value and leaves every other agent's value unchanged.
Since $M$ is finite, each agent has only finitely many possible bundle values, and thus only finitely many exchanges can occur while $S$ remains fixed.
\end{proof}

It remains to ensure that a suitable divider exists at every iteration. We next establish the packing lemmas used to do so.

\subsection{When Can a Divider Form Enough Bundles?}
\label{subsec:packing}

In this section, we find conditions under which a divider can form enough acceptable bundles for herself. Throughout this subsection, we assume that the divider's (non-negative, additive) valuation is given by the function $v$. For a threshold $t>0$, and a set $R\subseteq M$, we define its \emph{capped value} by
\[
    c_t(R)\coloneqq\sum_{g\in R}\min\{v(g),t\}.
\]
Capping limits the contribution of a single good to the value needed for one acceptable bundle. If $v(B)\ge t$, then $c_t(B)\ge t$: either $B$ contains a good worth at least $t$, or capping changes none of its values. 
Consequently, if $M$ contains $d$ pairwise disjoint bundles, each of value at least $t$, then
\begin{equation}
    c_t(M)\ge dt.
    \label{eq:capped-target}
\end{equation}
In particular, for agent $i$ and $0<t\le\mu_i^d$, a partition witnessing $\mu_i^d$ consists of $d$ bundles, each worth at least $t$. Thus, taking $v=v_i$, we obtain $c_t(M)\ge dt$.

\begin{lemma}[Greedy packing]
\label{lem:greedy-packing}
Let $u\ge1$ be an integer.
If the capped value $c_t(R)\ge(2u-1)t$, then $R$ contains $u$ pairwise disjoint bundles, each of value at least $t$.
\end{lemma}

\begin{proof}
Every good of value at least $t$ is assigned as a singleton bundle.
Among the remaining goods, a bundle is formed by repeatedly adding goods until their total value reaches $t$. We stop when $u$ bundles are formed.

Notice that every formed bundle has capped value strictly below $2t$: a large singleton contributes $t$, and every other bundle has value below $t$ before its last good (itself worth less than $t$) is added.
If fewer than $u$ bundles are produced, the remaining goods have total value below $t$.
If $p\le u-1$ are the number of completed bundles, then
\[
    c_t(R)<2pt+t\le(2u-1)t.
\]
This is a contradiction to $c_t(R)\ge (2u-1)t$; thus, at least $u$ bundles of value at least $t$ can be formed via this greedy packing process.
\end{proof}

Suppose $k<n$ pairwise disjoint bundles, each worth less than $t$, have already been assigned.
If the initial capped value is at least $dt$, the unassigned goods have capped value at least $(d-k)t$.
\Cref{lem:greedy-packing} therefore guarantees the existence of $n-k$ acceptable bundles among the remaining unassigned goods whenever
\begin{equation}
    d\ge2n-k-1.
    \label{eq:ordinary-packing-condition}
\end{equation}

In the previous lemma, we only distinguished between goods whose worth was at least $t$ (assigned as singletons) and those that were not (packed greedily till the bundle value was at least $t$). Next, we also consider ``intermediate goods'' whose value is between $t/2$ and $t$.

\begin{lemma}[Packing with few medium goods]
\label{lem:sparse-packing}
Normalize the instance such that the threshold $t=1$.
Let $r$ be the number of goods in $R$ whose values lie in $(1/2,1)$ and let $u\ge 1$ be an integer. There are $u$ pairwise disjoint bundles in $R$, each of value at least $1$, whenever
\begin{equation}
    c_1(R)\ge
    \max\left\{\frac{3u-1}{2},\,
                u+\frac{r-1}{2}\right\}.
    \label{eq:sparse-packing-condition}
\end{equation}
\end{lemma}

The proof assigns goods worth at least $1$ as singletons and uses goods worth at most $1/2$ to complete bundles containing one medium good. These bundles have value below $3/2$.
If the small goods run out, the remaining medium goods are paired. Counting the bundles formed in these two cases gives the two bounds in~\eqref{eq:sparse-packing-condition}.
The full proof appears in \Cref{sec:appendix-general-additive}.

We apply this bound to a divider who initially has few goods of value more than $\theta t$, where $\theta\le1/2$.
Removing assigned bundles reduces the available capped value, but also reduces the number of bundles she must form.
The following lemma shows that the required division remains possible throughout the procedure.

\begin{lemma}
\label{lem:sparse-after-removal}
Let $t>0$ and $0< \theta \le 1/2$. If $c_t(M)\ge dt$, $2d\ge3n-1$, and if at most $n-1$ goods have value greater than $\theta t$, after removing any $k<n$ pairwise disjoint bundles, each of value less than $t$, the remaining goods contain $n-k$ pairwise disjoint bundles, each of value at least $t$.
\end{lemma}

\begin{proof}
Let $R$ be the remaining goods and set $u=n-k$.
Each removed bundle has capped value below $t$, so $c_t(R)\ge(d-k)t$.
Let $r$ be the number of goods in $R$ with values in $(t/2,t)$.
Since $\theta\le1/2$, each such good has value greater than $\theta t$, and hence $r\le n-1$.

After dividing all values by $t$, we apply \Cref{lem:sparse-packing}. Its two required bounds follow from
\[
    d-k\ge\frac{3n-1}{2}-k
        =\frac{3u-1}{2}+\frac{k}{2}
        \ge\frac{3u-1}{2}
\]
and
\[
    d-k\ge u+\frac{n-1}{2}
        \ge u+\frac{r-1}{2}.
\]
Thus, $R$ contains $u=n-k$ disjoint bundles, each worth at least $t$ under the original valuation.
\end{proof}

\subsection{When \EFone{} Guarantees the Ordinal Target}
\label{subsec:automatic-satisfaction}

Some agents receive their ordinal targets in every complete \EFone{} allocation. Identifying them allows us to focus the construction on the remaining agents.

Consider an agent $i$ and a target $t>0$. For each other non-empty bundle, \EFone{} allows us to remove one good so that the remaining bundle has value at most $v_i(A_i)$. If only a few goods are of value more than $\theta t$ to her, the removed goods contribute only a limited amount to the total capped value. Since $t\le\mu_i^d$ implies that the initial capped value is at least $dt$, her own bundle must be sufficiently valuable.

\begin{lemma}
\label{lem:automatic-satisfaction}
For an agent $i$, a threshold $t>0$, and $0<\theta\le1$, let $q_i\coloneqq \bigl|\{g\in M:v_i(g)>\theta t\}\bigr|$. For every complete \EFone{} allocation $A$,
\begin{equation*}
    \sum_{g\in M}\min\{v_i(g),t\}
    \le
    n v_i(A_i)+(n-1)\theta t+(1-\theta)q_i t.
\end{equation*}
Consequently, for an integer $q_0\ge1$, assume that $t\le\mu_i^d$ and $d\ge n+(n-1)\theta+(1-\theta)(q_0-1).$
If $q_i<q_0$, then $v_i(A_i)\ge t$.
\end{lemma}

\begin{proof}
Let $x\coloneqq v_i(A_i)$, and let $c_t$ denote capped value with respect to $v_i$, that is, $c_t(R)=\sum_{g\in R}\min\{v_i(g),t\} $ for every $R\subseteq M.$
For each agent $j\neq i$ with $A_j\neq\varnothing$, since $A$ is \EFone{}, we can choose a good $g_j\in A_j$ such that $v_i(A_j\setminus\{g_j\})\le x.$
Let $T$ be the set of these chosen goods. Since there is at most one such good from every other agent's bundle, $|T|\le n-1$.
After removing $T$, agent $i$ values every remaining bundle at most $x$, including her own bundle.
Since capping does not increase value and the allocation is complete,
\[
    c_t(M\setminus T)\le nx.
\]

We next bound $c_t(T)$. Let $\ell$ be the number of goods in $T$ whose values to agent $i$ exceed $\theta t$; we know, $\ell\le q_i$.
Each of these $\ell$ goods has capped value at most $t$, while each of the other $|T|-\ell$ goods has capped value at most $\theta t$.
Therefore,
\[
    \begin{aligned}
    c_t(T)
    \le (|T|-\ell)\theta t+\ell t
    =|T|\theta t+(1-\theta)\ell t
    \le(n-1)\theta t+(1-\theta)q_i t.
    \end{aligned}
\]
Combining the bounds for $M\setminus T$ and $T$ gives
\[
    c_t(M)
    \le nx+(n-1)\theta t+(1-\theta)q_i t,
\] which proves the first claim.

For the second claim, suppose that $t\le\mu_i^d$ and $q_i<q_0$.
Since $t\le\mu_i^d$, there exists a partition of $M$ into $d$ bundles, each worth at least $t$ to agent $i$.
Each such bundle has capped value at least $t$; summing over the $d$ bundles gives $c_t(M)\ge dt$.
Moreover, since $q_i$ and $q_0$ are integers,
$q_i\le q_0-1$.

Towards a contradiction, suppose that $x<t$.
Then
\[
    \begin{aligned}
    dt
    &\le c_t(M)\\
    &\le nx+(n-1)\theta t+(1-\theta)q_i t\\
    &\le nx+(n-1)\theta t
             +(1-\theta)(q_0-1)t\\
    &<
    \bigl[n+(n-1)\theta
             +(1-\theta)(q_0-1)\bigr]t\\
    &\le dt.
    \end{aligned}
\]
This is a contradiction, so $v_i(A_i)\ge t$.
\end{proof}

We apply \Cref{lem:automatic-satisfaction} with $t=\mu_i^d$ for each agent whose target is positive.
Fix parameters $0<\theta\le1$ and an integer $q_0\ge1$ satisfying the conditions in the lemma, and define
\begin{equation}
    C\coloneqq
    \left\{
        i\in N:
        \mu_i^d>0,\ 
        \bigl|\{g\in M:v_i(g)>\theta\mu_i^d\}\bigr|
        \ge q_0
    \right\}.
    \label{eq:protected-agents}
\end{equation}
Every agent outside $C$ receives at least her ordinal target in any complete \EFone{} allocation.
Indeed, an agent with $\mu_i^d=0$ meets her target by non-negativity, whereas any other agent outside $C$ has fewer than $q_0$ goods of value more than $\theta\mu_i^d$, so the claim follows from \Cref{lem:automatic-satisfaction}.

Thus, it suffices to construct a complete \EFone{} allocation that explicitly guarantees $v_i(A_i)\ge\mu_i^d$ for every agent $i\in C$. The ordinal targets of all remaining agents then follow from \EFone{} and completeness.

\subsection{Reserving Goods as Anchors}
\label{subsec:anchors}

We now describe a construction that first reserves one good for each agent and then allocates additional goods to meet the remaining demands.
These goods are called \emph{anchors} and, for every agent $i\in C$, her anchor must be of value at least $\theta$-fraction of her target.

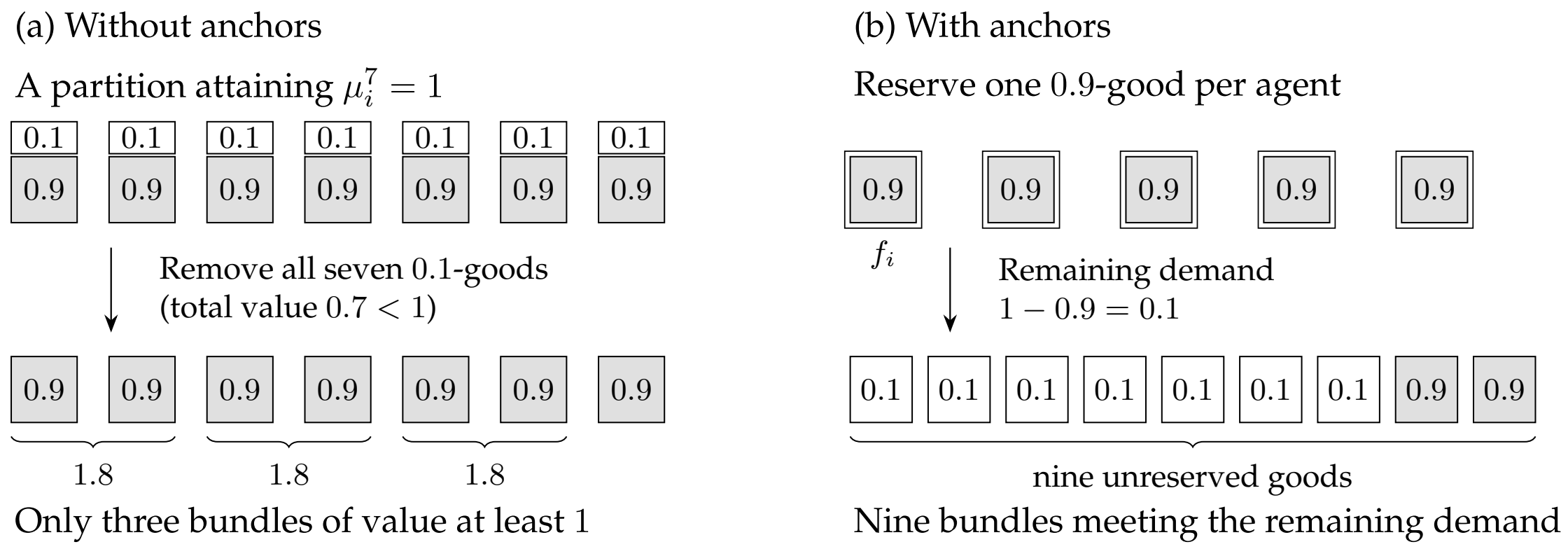
\begin{figure}[t]
    \centering
    \resizebox{0.9\linewidth}{!}{%
        \begin{tikzpicture}[x=1cm,y=1cm,font=\small,
  good/.style={draw=black,line width=.4pt},
  flow/.style={-{Stealth[length=1.8mm]},line width=.45pt}]

% Panel (a): the original partition and the goods left after removal.
\node[anchor=west] at (0,.55) {(a) Without anchors};
\node[anchor=west] at (0,-.05) {A partition attaining $\mu_i^7=1$};
\foreach \q in {0,...,6}{
  \pgfmathsetmacro{\xx}{.1+.98*\q}
  \filldraw[good,fill=black!12] (\xx,-1.4) rectangle ++(.66,.66);
  \node[font=\footnotesize] at (\xx+.33,-1.07) {$0.9$};
  \filldraw[good,fill=white] (\xx,-.71) rectangle ++(.66,.32);
  \node[font=\footnotesize] at (\xx+.33,-.55) {$0.1$};
}
\draw[flow] (1.1,-1.65)--(1.1,-2.5);
\node[anchor=west,align=left,font=\footnotesize] at (1.45,-2.07)
  {Remove all seven $0.1$-goods\\(total value $0.7<1$)};
\foreach \q in {0,...,6}{
  \pgfmathsetmacro{\xx}{.1+.98*\q}
  \filldraw[good,fill=black!12] (\xx,-3.4) rectangle ++(.66,.66);
  \node[font=\footnotesize] at (\xx+.33,-3.07) {$0.9$};
}
\foreach \q in {0,2,4}{
  \pgfmathsetmacro{\xx}{.1+.98*\q}
  \draw[line width=.4pt,decorate,
        decoration={brace,mirror,amplitude=3pt}]
    (\xx,-3.54)--(\xx+1.64,-3.54)
    node[midway,below=4pt,font=\footnotesize] {$1.8$};
}
\node[anchor=west] at (0,-4.42) {Only three bundles of value at least $1$};

% Panel (b): anchors and the remaining goods.
\begin{scope}[xshift=8.4cm]
\node[anchor=west] at (0,.55) {(b) With anchors};
\node[anchor=west] at (0,-.05) {Reserve one $0.9$-good per agent};
\foreach \q in {0,...,4}{
  \pgfmathsetmacro{\xx}{.1+1.38*\q}
  \filldraw[good,fill=black!12] (\xx,-1.4) rectangle ++(.66,.66);
  \draw[line width=.35pt] (\xx-.055,-1.455) rectangle ++(.77,.77);
  \node[font=\footnotesize] at (\xx+.33,-1.07) {$0.9$};
}
\node[font=\footnotesize] at (.43,-1.72) {$f_i$};
\draw[flow] (1.1,-1.65)--(1.1,-2.5);
\node[anchor=west,align=left,font=\footnotesize] at (1.45,-2.07)
  {Remaining demand\\$1-0.9=0.1$};
\foreach \q in {0,...,8}{
  \pgfmathsetmacro{\xx}{.1+.78*\q}
  \ifnum\q<7
    \filldraw[good,fill=white] (\xx,-3.4) rectangle ++(.62,.66);
    \node[font=\footnotesize] at (\xx+.31,-3.07) {$0.1$};
  \else
    \filldraw[good,fill=black!12] (\xx,-3.4) rectangle ++(.62,.66);
    \node[font=\footnotesize] at (\xx+.31,-3.07) {$0.9$};
  \fi
}
\draw[line width=.4pt,decorate,
      decoration={brace,mirror,amplitude=3pt}]
  (.1,-3.54)--(6.96,-3.54)
  node[midway,below=4pt,font=\footnotesize] {nine unreserved goods};
\node[anchor=west] at (0,-4.42) {Nine bundles meeting the remaining demand};
\end{scope}
\end{tikzpicture}%
    }
    \caption{The role of anchors. All values are from agent $i$'s
    perspective, with $n=5$, $m=14$, and $\mu_i^7=1$.
    (a) Removing a bundle worth $0.7$ leaves only three bundles
    worth at least $1$, while four agents still need bundles.
    (b) Reserving five anchors, shown with double borders,
    reduces $i$'s remaining demand to $0.1$; each of the nine
    unreserved goods meets this demand on its own.}
    \label{fig:anchor-intuition}
\end{figure}

To see why reserving a good helps, consider five agents and suppose that agent $i$ values seven goods at $0.9$ each and seven goods at $0.1$ each. Her $1$-out-of-$7$ maximin share is $1$. Assigning all seven $0.1$-goods to another agent removes only $0.7$ of value, but leaves enough goods for only three bundles worth at least $1$, while four agents still need bundles.
If we instead reserve a $0.9$-good for $i$, her remaining demand is only $0.1$.
\Cref{fig:anchor-intuition} illustrates how anchors can make it easier to form enough bundles meeting the remaining demand.

Throughout this subsection, assume that $|M|\ge n$ and that distinct goods can be assigned to the agents in $C$ so that each agent $i\in C$ receives a good of value at least $\theta\mu_i^d$.
We address the existence of such an assignment in \Cref{subsec:combining-constructions}.

\paragraph{Choosing the anchors.}
We reserve a distinct singleton good for every agent.
By assumption, we can first assign distinct goods to the agents in $C$ so that each agent $i\in C$ receives a good of value at least $\theta\mu_i^d$.
Since $|M|\ge n$, at least $n-|C|$ goods remain, so we can also give a distinct good to every agent outside $C$.

Among all assignments $(f_i)_{i\in N}$ of distinct goods satisfying $v_i(f_i)\ge\theta\mu_i^d$ for every $i\in C$, choose one that maximizes $\sum_{i\in N}v_i(f_i)$.
We call $f_i$ agent $i$'s \emph{anchor} and write $F\coloneqq\{f_i:i\in N\}$ to denote the set of all anchors.
This choice of $\{f_i\}_i$ provides two useful properties: 
(a) no agent prefers an unreserved good to her own anchor, and (b) the envy graph of the anchors contains no directed cycles.

\begin{lemma}
\label{lem:anchor-properties}
For every agent $i\in N$ and every good $g\in M\setminus F$, $v_i(g)\le v_i(f_i).$
Moreover, the directed graph $H$ on $N$, with an edge $i\to j$ whenever $v_i(f_j)>v_i(f_i)$, is acyclic.
\end{lemma}

\begin{proof}
Towards a contradiction, assume that some agent $i$ envies a good $g\in M\setminus F$ i.e. $v_i(g)>v_i(f_i)$.
Replacing $f_i$ by $g$ and leaving every other anchor unchanged gives a different assignment with a larger social welfare, while ensuring that, if $i\in C$, her new anchor still satisfies $v_i(g)\ge\theta\mu_i^d$. This contradicts the choice of anchors.

Next, again towards a contradiction, assume that $H$ contains a directed cycle $i_1\to i_2\to\cdots\to i_\ell\to i_1$.
Reassigning $f_{i_{r+1}}$ to agent $i_r$ for every $r\in[\ell]$, where $i_{\ell+1}=i_1$ gives each agent on the cycle a strictly more valuable anchor, while all other agents keep their anchors.
The sum of the agents' values for their anchors strictly increases, while ensuring that no agents value decreases (thus, the thresholds hold for agents in $C$), again contradicting maximality of the anchor assignment.
Hence, $H$ is acyclic.
\end{proof}

\paragraph{Allocating supplementary bundles.}
Keep the anchors reserved and apply \textsc{Shrink-and-Match} (\Cref{alg:divider}) from \Cref{subsec:divider} to the goods in $M\setminus F$, with the demands and divider selection rule specified below.
In this execution, $A_i$ denotes only the supplementary bundle assigned to agent $i$; it does not include $f_i$.
Set the demands to
\begin{equation}
    \tau_i=
    \begin{cases}
        \max\{0,\mu_i^d-v_i(f_i)\}, & i\in C,\\
        0, & i\notin C.
    \end{cases}
    \label{eq:anchor-demands}
\end{equation}
Thus, meeting the demand of an agent in $C$ ensures that her anchor and supplementary bundle together meet her ordinal target.

As in \textsc{Shrink-and-Match} (\Cref{alg:divider}), agents with zero demand are
satisfied from the start.
Initially, every supplementary bundle is empty and
$U=M\setminus F$.
Recall that $H$ is the envy graph of the reserved anchors: there is an edge $i\to j$ whenever $v_i(f_j)>v_i(f_i)$.
Let $H[N']$ denote its subgraph induced by the remaining agents $N'$.
The anchors remain fixed throughout the procedure, and $H$ always denotes their envy graph; supplementary bundles do not affect its edges.
At each iteration, we choose a sink of the induced graph $H[N']$ on the agents whose demands have not yet been met as the divider.
Such an agent always exists since $H$ is acyclic, and thus all subgraphs of $H$ must also be acyclic. 

If $i$ is the chosen divider, then
\[
    v_i(f_j)\le v_i(f_i)
    \qquad\text{for every }j\in N'.
\]
Consequently, every anchor that $i$ values more than her own belongs to an agent in $S$ (i.e. a satisfied agent).
This observation allows us to bound the value removed by the anchors when we verify that the requested division is possible.

\begin{lemma}[Packing after reserving anchors]
\label{lem:anchor-packing}
Let $0<\theta\le1$, $2d\ge3n-1$, and $d\ge2n-1-(n-1)\theta$.
With the anchor assignment and divider rule described above, every selected divider $i\in N'$ can form $|N'|$ pairwise disjoint bundles from $U$, each of value at least $\tau_i$ to her.
\end{lemma}

To see why the choice of divider helps, normalize her ordinal target to $1$. Let $a$ be the value of her anchor and $\delta=1-a$ be her remaining demand. If $k$ agents are already satisfied, at most $k$ anchors are worth more than $a$ to her. Every unreserved good is worth at most $a$, and each assigned supplementary bundle is worth less than $\delta$.

When $a\le2/3$, enough value remains to form the required bundles greedily after taking suitable singletons.
When $a>2/3$, the remaining demand is below $1/3$. Consider a partition into $d$ parts, each of value at least $1$ to the divider. Any part containing no anchor worth more than $a$ consists entirely of goods of value at most $a$.
Since the remaining demand is $\delta=1-a<1/3$, we can extract two disjoint bundles worth at least $\delta$ from each such part. We then bound how much capped value is lost when the anchors and previously assigned supplementary bundles are removed. The remaining capped value suffices to apply the greedy-packing lemma.
We refer the readers to \Cref{sec:appendix-general-additive} for the complete proof.

\paragraph{Appending the anchors.}
By \Cref{lem:anchor-packing}, every division requested by the procedure is possible. Hence, \Cref{lem:divider-guarantee} gives a possibly partial \EFX{} allocation $A$ of the non-anchor goods such that $v_i(A_i)\ge\tau_i$ for every agent $i$.

Give each agent her reserved anchor in addition to her supplementary bundle, and write
\[
    \widehat A_i\coloneqq A_i\cup\{f_i\}.
\]
The next lemma shows that the resulting allocation is \EFone{}.

\begin{lemma}
\label{lem:adjoining-anchors}
Let $(f_i)_{i\in N}$ be the anchors chosen above, and let $A$ be an \EFX{} allocation (possibly partial) of the goods in $M\setminus F$. Then the allocation defined by $\widehat A_i=A_i\cup\{f_i\}$ for every $i\in N$ satisfies \EFone{}.
\end{lemma}

\begin{proof}
Consider distinct agents $i,j\in N$.
We show that removing $j$'s anchor from $\widehat A_j$ eliminates any envy of $i$ toward $j$ i.e. $v_i(A_j)\le v_i(\widehat A_i)$.
If $A_j=\varnothing$, we are done; otherwise, choose any good $g\in A_j$.
Since the supplementary allocation $A$ is \EFX{}, $v_i(A_j\setminus\{g\})\le v_i(A_i).$
Moreover, $g$ is a non-anchor good, so \Cref{lem:anchor-properties} gives $v_i(g)\le v_i(f_i).$
Since valuations are additive, adding these inequalities, we get
\[
    \begin{aligned}
    v_i(\widehat A_j\setminus\{f_j\})
    =v_i(A_j)=v_i(A_j\setminus\{g\})+v_i(g)\le v_i(A_i)+v_i(f_i)=v_i(\widehat A_i).
    \end{aligned}
\]
Thus, $\widehat A$ is \EFone{}.
\end{proof}

The resulting allocation also meets the ordinal target of every agent in $C$.
That is, for each $i\in C$,
\[
    \begin{aligned}
    v_i(\widehat A_i)
    &=v_i(A_i)+v_i(f_i)\\
    &\ge\tau_i+v_i(f_i)\\
    &=\max\{0,\mu_i^d-v_i(f_i)\}+v_i(f_i)\\
    &\ge\mu_i^d.
    \end{aligned}
\]
Therefore, whenever the required anchor assignment exists and the anchor conditions in the statement of \Cref{lem:anchor-packing} hold, this construction gives a possibly partial \EFone{} allocation meeting the targets of all agents in $C$.
In \Cref{subsec:combining-constructions}, we complete the allocation while preserving \EFone{} and without decreasing any agent's value.
The targets of agents outside $C$ then follow from \Cref{lem:automatic-satisfaction}.

\subsection{The Construction With and Without Anchors}
\label{subsec:combining-constructions}

We use the following completion property of envy-cycle elimination~\citep{LMM+04approximately}; we include the proof in \Cref{sec:appendix-general-additive} for completeness.

\begin{lemma}
\label{lem:ef1-completion}
Any partial \EFone{} allocation can be completed while preserving \EFone{} and without decreasing any agent's value.
\end{lemma}

We now combine the two constructions. If a matching reserves suitable anchors for every agent in $C$, we use \Cref{subsec:anchors}. Otherwise, Hall's theorem identifies a group of agents with few valuable goods who can serve as dividers in the early iterations.

The following conditions ensure that both branches succeed. The first two support the packing arguments, the third guarantees the targets of agents outside $C$, and the fourth ensures that enough suitable dividers remain available until ordinary greedy packing can be applied.

\begin{proposition}
\label{prop:ordinal-ef1-parameters}
Let $n\ge2$ and $d\ge n$ be integers.
Suppose there exist $0<\theta\le1/2$ and an integer $q_0\ge1$ such that
\begin{align}
    2d&\ge3n-1,
    \label{eq:parameter-sparse}\\
    d&\ge2n-1-(n-1)\theta,
    \label{eq:parameter-anchor}\\
    d&\ge n+(n-1)\theta+(1-\theta)(q_0-1),
    \label{eq:parameter-automatic}\\
    q_0+1&\ge2n-d-1.
    \label{eq:parameter-dividers}
\end{align}
Then every additive instance admits a complete allocation that is both \EFone{} and ($1$-out-of-$d$)-\MMS{}.
\end{proposition}

\Cref{alg:additive-construction} summarizes the construction.
The matching determines which of two approaches we use: either reserve anchors and meet the remaining demands using the other goods, or use the agents identified by Hall's theorem as dividers in the early iterations.
In both cases, we finish by completing the allocation through
envy-cycle elimination.
Within each call to \Cref{alg:divider}, $S$ denotes the set of satisfied agents i.e. those whose demands have been met.
Since the construction uses the exact ordinal targets $\mu_i^d$, even though \Cref{lem:divider-guarantee} guarantees finite termination, we do not establish a polynomial running-time bound for the construction.

\begin{algorithm}[t]
\caption{Construction for general additive valuations}
\label{alg:additive-construction}
\begin{algorithmic}[1]
\REQUIRE An additive instance; parameters $d,\theta,q_0$ satisfying \Cref{prop:ordinal-ef1-parameters}; targets $t_i=\mu_i^d$
\ENSURE A complete \EFone{} and $1$-out-of-$d$-\MMS{} allocation

\IF{$t_i=0$ for every $i\in N$}
    \STATE Complete the empty allocation by envy-cycle elimination
    \STATE \textbf{return} the resulting allocation
\ENDIF

\STATE $C\gets \{i\in N:t_i>0,\ |\{g\in M:v_i(g)>\theta t_i\}|\ge q_0\}$
\STATE Construct a bipartite graph $G$ between $C$ and $M$, with edge $(i,g)$ when $v_i(g)\ge\theta t_i$

\IF{$G$ has a matching covering $C$}
    \STATE Choose distinct anchors $(f_i)_{i\in N}$ maximizing $\sum_{i\in N}v_i(f_i)$, subject to $v_i(f_i)\ge\theta t_i$ for $i\in C$
    \STATE $F\gets\{f_i:i\in N\}$
    \STATE Set $\tau_i\gets\max\{0,t_i-v_i(f_i)\}$ for $i\in C$, and $\tau_i\gets0$ otherwise
    \STATE Construct $H$ with edge $i\to j$ when $v_i(f_j)>v_i(f_i)$
    \STATE Run \Cref{alg:divider} on $M\setminus F$ with demands $(\tau_i)_{i\in N}$:
    \STATE \hspace{\algorithmicindent}
           choose a sink of $H[N\setminus S]$ as divider
    \STATE \hspace{\algorithmicindent}
           form bundles as in \Cref{lem:anchor-packing}
    \STATE Let $A$ be the returned supplementary allocation
    \STATE $A_i\gets A_i\cup\{f_i\}$ for every $i\in N$
\ELSE
    \STATE Choose $X\subseteq C$ with $|\Gamma_G(X)|<|X|$
    \STATE $K\gets2n-d-1$
    \STATE Run \Cref{alg:divider} on $M$ with demands $\tau_i=t_i$:
    \STATE \hspace{\algorithmicindent}
           while $|S|<K$, choose a divider from $X\setminus S$ and use \Cref{lem:sparse-after-removal}
    \STATE \hspace{\algorithmicindent}
           otherwise, choose any remaining agent and use \Cref{lem:greedy-packing}
    \STATE Let $A$ be the returned allocation
\ENDIF

\STATE Complete $A$ by envy-cycle elimination
\STATE \textbf{return} $A$
\end{algorithmic}
\end{algorithm}

\begin{proof}
If every target $\mu_i^d$ is zero, we can simply apply \Cref{lem:ef1-completion} to the empty allocation.
Otherwise, at least one agent has a positive target, and a partition witnessing that target has $d$ non-empty parts, so $|M|\ge d\ge n$.

Define $C$ as in~\eqref{eq:protected-agents}.
By~\eqref{eq:parameter-automatic} and \Cref{lem:automatic-satisfaction}, every agent outside $C$ meets her target in any complete \EFone{} allocation.
Construct a bipartite graph between $C$ and $M$, placing an edge between $i$ and $g$ whenever $v_i(g)\ge\theta\mu_i^d$.

\vspace{0.5em}

\underline{\emph{Case 1: a matching covers $C$.}}
Since $|M|\ge n$, extend this matching to an assignment of distinct anchors to all agents and choose the maximizing assignment described in \Cref{subsec:anchors}.
Run \textsc{Shrink-and-Match} (\Cref{alg:divider}) on the non-anchor goods with demands \eqref{eq:anchor-demands}, choosing a sink of $H[N']$ as each divider.
Conditions~\eqref{eq:parameter-sparse} and~\eqref{eq:parameter-anchor}, together with \Cref{lem:anchor-packing}, ensure that every requested division is possible.
By \Cref{lem:divider-guarantee}, the procedure returns supplementary bundles satisfying \EFX{} and all these demands.

Append the anchors to this allocation.
Every agent in $C$ now receives value at least $\mu_i^d$, and \Cref{lem:adjoining-anchors} gives \EFone{}.
Complete the allocation using \Cref{lem:ef1-completion}.
The targets of agents in $C$ remain satisfied, and the targets of agents outside $C$ follow from \Cref{lem:automatic-satisfaction}.

\vspace{0.5em}
\underline{\emph{Case 2: no matching covers $C$.}}
Hall's theorem gives a non-empty $X\subseteq C$ with $|\Gamma(X)|<|X|$.
Every agent in $X$ has at least $q_0$ neighbors, so
\[
    |X|\ge q_0+1.
\]
Also, every good worth more than $\theta\mu_i^d$ to an agent $i\in X$ belongs to $\Gamma(X)$.
Thus, each such agent has at most $|\Gamma(X)|\le n-1$ goods above this threshold.

Run \textsc{Shrink-and-Match} (\Cref{alg:divider}) on all goods $M$, with $\tau_i=\mu_i^d$ for every $i\in N$.
Set $K=2n-d-1$.
While $k=|S|<K$, choose a divider from $X\setminus S$.
Such an agent exists because
\[
    |X|\ge q_0+1\ge K>k.
\]
For this divider, the target partition gives capped value at least $d$ after normalization, and every assigned bundle is worth below her target by~\eqref{eq:unassigned-demand}.
\Cref{lem:sparse-after-removal} therefore supplies the requested division.

Once $k\ge K$, any remaining agent can serve as the divider: the inequality $k\ge2n-d-1$ is equivalent to $d\ge2n-k-1$, so ordinary greedy packing applies by \eqref{eq:ordinary-packing-condition}.
If $K\le0$, use this rule from the outset.
\Cref{lem:divider-guarantee} gives an \EFX{} allocation meeting every target.
Complete it using \Cref{lem:ef1-completion}, preserving all targets and \EFone{}.
\end{proof}

\subsection{Choosing the Parameters for the Golden-Ratio Bound}
\label{subsec:ordinal-guarantees}

We now choose parameters satisfying the conditions of \Cref{prop:ordinal-ef1-parameters} that give the golden-ratio ordinal \MMS{} guarantee.

\paragraph{Deriving the parameter choice.}
Let $K\coloneqq2n-d-1$.
As shown in the proof of \Cref{prop:ordinal-ef1-parameters}, ordinary greedy packing suffices once at least $K$ agents are satisfied.
Equivalently, $d=2n-K-1$, so obtaining a smaller denominator is equivalent to choosing a larger value of $K$.
In terms of $K$, the anchor condition
\eqref{eq:parameter-anchor} becomes
\[
    2n-K-1\ge2n-1-(n-1)\theta,
\]
or equivalently, $\theta\ge K/(n-1)$.
Similarly, the divider condition \eqref{eq:parameter-dividers} requires $q_0\ge K-1$.
We choose both parameters at these lower bounds:
\begin{equation}
    \theta=\frac{K}{n-1},
    \qquad
    q_0=K-1.
    \label{eq:chosen-parameters}
\end{equation}
These choices satisfy the anchor and divider conditions with equality. To satisfy $q_0\ge1$ and $0<\theta\le1/2$, we require $2\le K\le\frac{n-1}{2}$; the upper bound also guarantees that \eqref{eq:parameter-sparse} holds. 
Moreover, $K$ must be an integer because $d=2n-K-1$ is an integer.
This range is non-empty precisely when $n\ge5$, which is why we handle smaller values of $n$ separately.

It remains to satisfy the automatic-satisfaction condition \eqref{eq:parameter-automatic}.
Substituting $d=2n-K-1$ and the choices in~\eqref{eq:chosen-parameters}, this condition becomes
\[
    2n-K-1
    \ge
    n+K+\left(1-\frac{K}{n-1}\right)(K-2).
\]
Multiplying by $n-1$ and rearranging gives
\begin{equation}
    K^2-(3n-1)K+n^2-1\ge0.
    \label{eq:parameter-quadratic}
\end{equation}

The smaller root of this quadratic is
\[
    \kappa(n)
    \coloneqq
    \frac{3n-1-\sqrt{5n^2-6n+5}}{2}.
\]

Since the quadratic has positive leading coefficient, \eqref{eq:parameter-quadratic} holds when $K$ is at most the smaller root or at least the larger root.
Even though we want $K$ to be as large as possible, the latter possibility is excluded since the larger root is more than $(n-1)/2$, which is our upper bound on $K$.
Thus, we require $K\le\kappa(n)$.
To minimize $d=2n-K-1$, we choose the largest integer satisfying this inequality, namely $K=\lfloor\kappa(n)\rfloor$ which gives
\[
    d=2n-1-\lfloor\kappa(n)\rfloor
    =
    \left\lceil
        \frac{n-1+\sqrt{5n^2-6n+5}}{2}
    \right\rceil.
\]

This choice gives the denominator in the following theorem.
The proof in \Cref{sec:appendix-general-additive} verifies the parameter restrictions, handles $n\le4$ separately, and compares the denominator with $\lfloor\varphi n\rfloor$.

\begin{theorem}
\label{thm:golden-ratio-ef1}
Every additive instance admits a complete allocation that is both \EFone{} and ($1$-out-of-$D(n)$)-\MMS{}, where
\[
    D(n)
    \coloneqq
    \left\lceil
        \frac{n-1+\sqrt{5n^2-6n+5}}{2}
    \right\rceil.
\]
Moreover, $D(n) \le\lfloor\varphi n\rfloor $, where $\varphi=(1+\sqrt5)/2$ denotes the golden ratio. Thus, every additive instance admits an allocation that is simultaneously \EFone{} and ($1$-out-of-$\lfloor\varphi n\rfloor$)-\MMS{}.
\end{theorem}

For some values of $n$, $D(n)$ is strictly smaller than $\lfloor\varphi n\rfloor$; for example, $D(5)=7<8=\lfloor5\varphi\rfloor$.

\section{Bivalued Valuations}
\label{sec:bv}

For bivalued valuations, each agent $i\in N$ values every good at either $a_i$ or $b_i$, where $0\le a_i\le b_i$.
The two values, as well as the goods with a larger value, may differ across agents. In this section, we show that this additional structure on valuations allows us to obtain stronger simultaneous guarantees as well as a polynomial-time algorithm.

\begin{theorem}
\label{thm:bv-main}
Every instance with bivalued valuations admits a complete allocation that is simultaneously \EFX{} and $(1\text{-out-of-}\left\lceil\frac{7n-1}{6}\right\rceil)$-\MMS{}.
If the values of the goods are rational numbers represented in binary, such an allocation can be computed in polynomial time.
\end{theorem}

\paragraph{Overview of the construction and the analysis.}

An algorithm (\MatchAndPause{}; \Cref{alg:bivalued}) is first given for $0< a_i < b_i$ and is then extended for zero and constant values.
Here, we allocate goods in rounds, giving each agent at most one good in a round. We first choose a matching that assigns as many agents as possible a good they value highly. If an unmatched agent could obtain a high-valued good by reassigning goods along an alternating path, we temporarily pause the matched agents reached by such paths.
The unmatched agent continues receiving goods during these pauses.
Among maximum-cardinality matchings, we favor agents with larger high-to-low value ratios. This choice helps us bound the length of each pause in terms of the paused agent's own ratio.

To analyze the allocation from agent $i$'s perspective, we look at the last round in which a good she values highly is assigned. After that round, every remaining good is low for her. Whether she receives a high good in that round, and which agents are then paused, lets us bound the numbers of high and low goods in the other bundles. These bounds show that removing any good from a bundle she envies eliminates her envy. They also give us a way to prove the ordinal guarantee.
When the value of her own bundle is large enough, a bound on the total value of the goods suffices. For smaller values, we assign auxiliary weights to goods and show that the total weight is too small to support $d$ disjoint bundles that the agent strictly prefers to her own bundle.
These weights are used only in the analysis; the algorithm does not compute the agents' maximin shares or the auxiliary weights.

\subsection{Algorithm for Bivalued Goods}
\label{subsec:bv-algorithm}

We first consider valuations with $0<a_i<b_i$ for every agent i.e. each agent has one of two distinct positive values for goods. We will extend the analysis to non-negative values, with some agents possibly having constant values across all goods at the end of the section. 

Dividing agent $i$'s values by $a_i$ preserves her envy comparisons and scales each of her maximin shares by the same factor. Thus, we can assume that
\[
    v_i(g)\in\{1,r_i\},
    \qquad
    r_i\coloneqq b_i/a_i>1.
\]
A good is \emph{high} for agent $i$ if it has value $r_i$ to her, and is \emph{low} for her otherwise.

\begin{algorithm}[t]
\caption{\MatchAndPause{} for positive bivalued valuations}
\label{alg:bivalued}
\begin{algorithmic}[1]
\REQUIRE Bivalued valuations $v_i(g)\in\{1,r_i\}$ with $r_i>1$
\ENSURE A complete allocation that is \EFX{} and ($1$-out-of-$\lceil \frac{7n-1}{6}\rceil$)-\MMS{}

\STATE All goods are unallocated $A_i\gets\varnothing$ for every $i\in N$; all agents are unmarked; every $i\in N$ is active $\rho_i \gets 1$; $t\gets 1$

\WHILE{unallocated goods remain}
    \PHASE{Match active agents to high-value goods}
    \STATE Let $P$ be the agents active in round $t$ i.e. $P \gets \{i\in N : \rho_i \le t\}$. Construct bipartite graph $G$ between $P$ and the unallocated goods, with an edge $(i,g)$ whenever {$v_i(g)=r_i$}
    \STATE Compute a maximum-cardinality matching $Q$ in $G$; among all such matchings, choose one maximizing $\sum r_i$ summed over matched agents \label{line:matching}
    \STATE Let $T\subseteq P$ be the agents left unmatched
    \PHASE{Mark and pause agents reached by alternating paths}
    \STATE Direct the edges of $G$ outside $Q$ from agents to goods and the edges in $Q$ from goods to agents
    \STATE For each matched agent $j$, let $R_j$ be the set of agents in $T$ from which $j$ is reachable in this directed graph
    \STATE For every matched agent $j$ with $R_j\neq\varnothing$, mark $j$ in round $t$ and pause her for the next $s_j\coloneqq\max_{u\in R_j}\lfloor r_u\rfloor-1$ rounds; set $\rho_j \gets t+s_j+1$ \label{line:pause}
    \PHASE{Allocate}
    \STATE Give every matched agent her matched good
    \STATE Give each agent in $T$ one remaining good (worth $1$ to her) until goods run out, serving never-marked agents first, then marked agents in order of the round in which they were marked, breaking ties by the fixed agent order\label{line:fill}
    \STATE $t\gets t+1$
\ENDWHILE
\RETURN Allocation $(A_i)_{i\in N}$
\end{algorithmic}
\end{algorithm}

\Cref{alg:bivalued} describes the procedure. It distinguishes between pausing an agent and marking her. An active agent participates in the current round; a paused agent skips rounds until her pause expires. An agent marked in round $t$ with pause length $s_j$ skips rounds $t+1,\ldots,t+s_j$ and returns in round $t+s_j+1$. In particular, a pause of length zero does not cause her to miss a round. For each agent $i$, we use $\rho_i$ to record the earliest round in which she may next participate.

On the other hand, a mark is permanent and is used to determine the priority among agents while assigning goods to unmatched agents. Never-marked agents are served first, followed by marked agents in order of their marking round. We fix an arbitrary agent order to break any remaining ties.

\begin{figure}[t]
\centering
\begin{tikzpicture}[
    >={Stealth[length=2.2mm]},
    agent/.style={circle, draw, thick, minimum size=7mm, inner sep=0pt},
    unmatched/.style={agent, double, double distance=1.2pt},
    paused/.style={agent, fill=gray!25},
    good/.style={rectangle, draw, thick, minimum size=6mm, inner sep=1pt},
    rest/.style={good, draw=gray, text=gray},
    match/.style={->, line width=1.8pt},
    nonmatch/.style={->, thin, dashed},
    lab/.style={font=\small},
]
  % agents
  \node[unmatched] (a1) at (0, 0)    {$1$};
  \node[paused]    (a2) at (0,-1.5)  {$2$};
  \node[paused]    (a3) at (0,-3)    {$3$};
  \node[lab, left=3mm of a1] {$r_1=3$};
  \node[lab, left=3mm of a2] {$r_2=4$};
  \node[lab, left=3mm of a3] {$r_3=5$};
  % goods
  \node[good] (g1) at (3.6, 0)     {$g_1$};
  \node[good] (g2) at (3.6,-0.75)  {$g_2$};
  \node[rest] (g3) at (3.6,-1.5)   {$g_3$};
  \node[rest] (g4) at (3.6,-2.25)  {$g_4$};
  \node[rest] (g5) at (3.6,-3)     {$g_5$};
  % column labels
  \node[lab, above=2mm of a1] {agents};
  \node[lab, above=2mm of g1] {goods};
  % non-matching high-value edges: agent -> good
  \draw[nonmatch] (a1) -- (g1);
  \draw[nonmatch] (a2) -- (g2);
  % matching edges: good -> agent
  \draw[match] (g1) -- (a2);
  \draw[match] (g2) -- (a3);
\end{tikzpicture}
\caption{An execution of \MatchAndPause{}.
There are edges from agents to her high-valued goods. The matching $Q=\{(2,g_1),(3,g_2)\}$ is denoted by the thick edges; it maximizes the sum of matched agents' ratios among maximum-cardinality matchings.
The unmatched agent~$1$ reaches both agents~$2$ and~$3$ along $1\to g_1\to 2\to g_2\to 3$.
Agents $2$ and $3$ receive their matched goods ($g_1, g_2$, respectively) but are paused for $\lfloor r_1\rfloor-1=2$ rounds.
Agent~$1$ receives $g_3,g_4,g_5$ in successive rounds, reaching value $3$, equal to her value for agent~$2$'s bundle.
Notice that agent~$3$ is paused even though $g_2$ is low for agent~$1$.}
\label{fig:bivalued-example}
\end{figure}
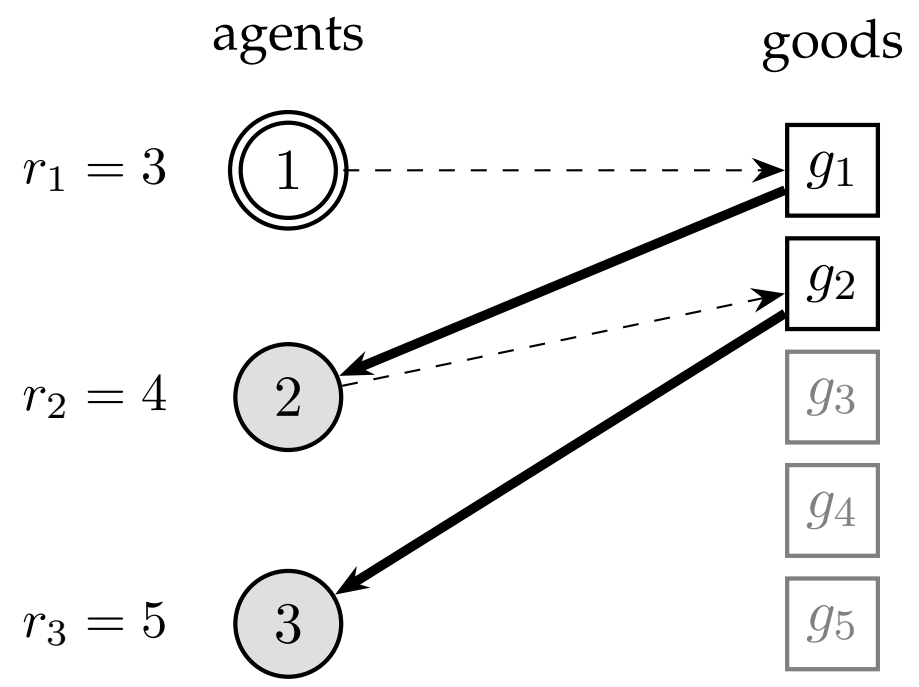

\Cref{fig:bivalued-example} illustrates why marking follows alternating paths: an unmatched agent may cause a matched agent to be paused even when she does not value that agent's good highly.

The matching rule ensures that, once an agent is marked, all remaining goods are low for her. Thus, she can be marked only once. The preference for larger ratios among maximum-cardinality matchings also ensures that her pause is not too long relative to her own ratio.

\begin{lemma}
\label{lem:bv-reachability}
The allocation procedure has the following properties.
\begin{enumerate}
    \item[(i)]
    An unmatched active agent, and every matched agent reached from an unmatched active agent, has no high good among the goods remaining after the current round.

    \item[(ii)]
    Each agent is marked at most once.

    \item[(iii)]
    If an unmatched agent $u$ reaches a matched agent $j$, then $r_u\le r_j$. Consequently,
    $s_j\le\lfloor r_j\rfloor-1$.

    \item[(iv)]
    The procedure allocates all goods in at most $m$ rounds and can be implemented in polynomial time for rational good values.
\end{enumerate}
\end{lemma}

\begin{proof}
Consider a fixed round and the matching $Q$ that happens in this round. If an unmatched good were reachable from an unmatched agent, a shortest such path would be an augmenting path for $Q$, contradicting that the matching selected was the maximum cardinality one. Thus, every good reachable from an unmatched agent is matched. Every high good of an unmatched agent is reachable. The same holds for every high good of a reached matched agent: her matching good is already on a path that can be reached, and each of her other high goods can be reached by one more edge. All these goods are assigned in the current round, proving (i).

An agent is marked only when she is matched and reached from an unmatched agent. By (i), after that round she has no high good remaining. She can therefore never again be matched in a high-edge graph, so she cannot be marked again. This proves (ii).

Consider an unmatched agent $u$ that reaches a matched agent $j$. Choose a simple alternating path from $u$ to $j$ and exchange its matching and non-matching edges. The resulting matching has the same cardinality as $Q$. The internal agents remain matched, $u$ becomes matched, and $j$ becomes unmatched. The choice of $Q$ therefore implies $r_u\le r_j$. Taking the maximum over all unmatched agents reaching $j$ proves the bound on $s_j$, establishing (iii).

Before any pause occurs, every agent is active. If some agent is paused, her marking round had an unmatched agent $u$ with an alternating path to a high good. Property (i) implies that $u$ has no high good remaining after this round. 
Since $u$ is unmatched, she is not marked or paused in this round. Having no high good remaining, she can never be matched or paused in a later round. She therefore remains active until the procedure ends. 
Thus, at least one agent is active whenever goods remain. Every round therefore assigns at least one good, so there are at most $m$ rounds.

Each round uses a bipartite matching computation and graph searches.
For example, the two matching objectives can be implemented by giving each high edge $(i,g)$ weight $B+r_i$, where $B=n\max_i r_i+1$, and computing a maximum-weight matching.
Increasing cardinality by one improves the $B$ contribution by more than the possible change in the sum of ratios. Among matchings of equal cardinality, the weights maximize that sum. Since all these weights have polynomial bit length, (iv) holds.
\end{proof}

\subsection{Structural Lemmas and the Envy Guarantee}
\label{subsec:bv-structure}

Consider an agent $i$ and write $r\coloneqq r_i$. Let $h$ and $\ell$ denote the numbers of high and low goods in her final bundle $A_i$, respectively, and thus, we have $x\coloneqq v_i(A_i)=hr+\ell.$
For another bundle $A_j$, let $H$ and $L$ denote the numbers of high and low goods according to this same agent $i$ i.e. $v_i(A_j)=Hr+L$.

In the following lemma, we present a characterization of the ordinal guarantee that will be useful throughout the analysis.

\begin{lemma}
\label{lem:bv-strict-packing}
For any agent $i$, integer $d\ge n$, and $x\ge0$, we have $x\ge\mu_i^d$ if and only if there do not exist $d$ pairwise disjoint bundles $B_1,\ldots,B_d\subseteq M$ such that $v_i(B_j)>x$ for every $j\in[d]$.
\end{lemma}

\begin{proof}
If $\mu_i^d>x$, then the partition attaining $\mu_i^d$ consists of $d$ bundles of value strictly greater than $x$. Conversely, if there are $d$ disjoint bundles of value strictly greater than $x$, distributing any unallocated goods arbitrarily among these bundles produces a partition whose minimum value is larger than $x$ i.e. $\mu_i^d>x$.
\end{proof}

The next lemma describes the output allocation from each agent’s perspective. For each agent, at least one of two conditions must hold. Condition (a) bounds the value of every other bundle and restricts the number of high and low goods it can contain. Condition (b) guarantees that the agent receives at least her maximin share.

\begin{lemma}
\label{lem:bv-structure}
Assume that agent $i$ has at least one high good in $M$. The output allocation satisfies at least one of the following conditions.
\begin{enumerate}
    \item[\textnormal{(a)}]
    Every other bundle satisfies
    \[
        H\le h+1,
        \qquad
        Hr+L\le x+1.
    \]
    Moreover, if $H=h+1$, then $L\le\max\{0,\ell-\lfloor r\rfloor\}.$

    \item[\textnormal{(b)}]
    Agent $i$ receives her maximin share: $x\ge\mu_i$.
\end{enumerate}
In every case, agent $i$ does not \EFX{}-envy any other agent's bundle.
If no good is high for $i$, then $x\ge\mu_i$, and agent $i$ does not \EFX{}-envy any other agent's bundle.
\end{lemma}

\begin{proof}
First, we consider the case where at least one good is high for agent $i$. Let $t$ be the last round in which any good high for $i$ is assigned, and let $R$ be the final round. Before round $t$, agent $i$ is active, unmarked, and receives one high good in every round (holds by \Cref{lem:bv-reachability}(i)).
Every agent receives at most one good in each round, and every active agent receives a good except possibly in the final round.

\paragraph{Case 1: Agent $i$ receives a high good in round $t$.}
We have $h=t$. Agent $i$ is matched in round $t$: if she were unmatched, all her high goods would be matched to other agents in that round (see the proof of \Cref{lem:bv-reachability}(i)), so the good she receives would be low for her. Since no high good for $i$ is allocated after round $t$, every other bundle has at most $h$ such goods. Let $s$ be the pause assigned to $i$ in round $t$, with $s=0$ if she is not marked. By \Cref{lem:bv-reachability}, $0\le s\le\lfloor r\rfloor-1.$
After this pause, $i$ remains active and receives only low goods.

Consider an agent $j$ who does not receive an $i$-high good in round $t$. Her value according to $i$ is at most $v_i(A_j)\le(t-1)r+1+(R-t)$.

If $R-t\ge s$, then agent $i$ is active in the $R-t-s$ rounds following her pause. She receives a low good in each of these rounds, except possibly in the final round. Together with the $t$ high goods received through round $t$, this gives $x\ge tr+(R-t-s)-1$.
Consequently,
\[
    v_i(A_j)-x
       \le s+2-r
       \le\lfloor r\rfloor+1-r
       \le1,
\]
where we use $s\le \lfloor r \rfloor - 1$ (\Cref{lem:bv-reachability}(iii)).

If $R-t<s$, agent $i$ receives no good after round $t$, and $x=tr$.
In that case,
\[
    v_i(A_j)-x\le R-t+1-r< s + 1 -r  \le \lfloor r \rfloor - r \le 0 < 1.
\]

Now consider an agent $j$ who does receive a high good according to agent $i$ in round $t$. If $i$ is marked, every unmatched agent that reaches $i$ also reaches $j$, by extending its alternating path along the non-matching edge from $i$ to $j$'s matched good. Hence $s_j\ge s$, and $j$ is paused for at least $s$ rounds. If $i$ is unmarked, $s=0$ and the bounds below hold nonetheless. In either event,
\[
    v_i(A_j)\le tr+\max\{0,R-t-s\},
    \qquad
    x\ge tr+\max\{0,R-t-s\}-1.
\]
Thus $v_i(A_j)\le x+1$ for every $j\ne i$. Since every other bundle has $H\le h$, no bundle has $H=h+1$, so the additional restriction in condition~(a) is automatically satisfied. This proves condition~(a).

\paragraph{Case 2: Agent $i$ does not receive a high good in round $t$.}
Agent $i$ is unmatched in round $t$. She has not been marked earlier, and she can never be marked later because no high good for her remains. Hence she is active in every round. Let $k\coloneqq\lfloor r\rfloor$ and let $\varepsilon\in\{0,1\}$ indicate whether $i$ misses a good in the final round. We have $h=t-1$. Then
\begin{equation}
    \ell=R-h-\varepsilon,
    \qquad
    R-t=\ell+\varepsilon-1.
    \label{eq:bv-round-count}
\end{equation}
Every other bundle contains at most $R\le h+\ell+1$ goods and at most $t=h+1$ goods high for $i$.

Every agent $j$ who receives an $i$-high good $g$ in round $t$ is matched in that round and is reachable from $i$ along the directed path $i\to g\to j$.
Thus she is marked and paused for at least $k-1$ rounds i.e. $s_j \ge k-1$. If she subsequently returns, she has no good high for herself remaining, by \Cref{lem:bv-reachability}(i), and so is unmatched in every later round. If $i$ misses a good in the final round, $j$ must miss one as well: the never-marked agent $i$ is processed before the marked agent $j$. Therefore, if $j$ returns before the execution ends, the number of goods she receives after round $t$ is at most
\[
    R-t-s_j-\varepsilon = \ell + \varepsilon -1 -s_j -\varepsilon
       \le\ell-k.
\]
All these goods are low for $i$. If $j$ does not return, she receives no subsequent good. A bundle with $H=h+1$ must contain a high good in every round through $t$, and hence has no earlier low good.
It follows that
\begin{equation}
    H=h+1
    \quad\Longrightarrow\quad
    L\le\max\{0,\ell-k\}.
    \label{eq:bv-extra-high}
\end{equation}

An agent who does not receive an $i$-high good in round $t$ has value at most
\[
    hr+1+(R-t)=x+\varepsilon\le x+1.
\]
If an agent receives an $i$-high good in round $t$ and subsequently returns, then $R-t\ge s_j+1\ge k$.
Since $R-t=\ell+\varepsilon-1\le\ell$, this implies $\ell\ge k$.
Her bundle has value at most $(h+1)r+\ell-k$, and hence exceeds $x$ by at most $r-k<1$. 
An agent who receives such a good and remains paused has value at most $(h+1)r$, and hence exceeds $x$ by at most $r-\ell$. If $\ell+1\ge r$, all these bounds are at most $x+1$. Together with $H\le h+1$ and \eqref{eq:bv-extra-high}, this proves condition~(a).

Finally, consider the case where $\ell+1<r$.
We will prove condition (b) (i.e. $x\ge \mu_i$) by showing that there are too few goods to form $n$ disjoint bundles that $i$ strictly prefers to $A_i$. 

First, consider the bundles produced by the algorithm. Since $\ell+1$ is an integer and $\ell+1<r$, we have $\ell+1\le\lfloor r\rfloor=k$. By \eqref{eq:bv-round-count}, $R-t=\ell+\varepsilon-1\le\ell\le k-1$. Thus, at most $k-1$ rounds remain after round $t$. Every agent who receives an $i$-high good in round $t$ is paused for all remaining rounds and receives no further goods. Her bundle therefore contains at most $t=h+1$ goods. For any other agent, all $i$-high goods in her bundle were received in the first $t-1=h$ rounds. Her bundle thus contains at most $h$ high goods and at most $R=h+\ell+\varepsilon\le h+\ell+1$ goods in total.

To use these counts, assign an auxiliary weight of $\ell+1$ to each $i$-high good and a weight of $1$ to each low good. 
With this choice, one high good weighs more than all $\ell$ low goods in $A_i$ together, while its weight remains below its value $r$.

Let $w(B)$ denote the total weight of a bundle $B$, and define $W\coloneqq(h+1)(\ell+1)$ as the weight of $h+1$ high goods. The weight of $A_i$ is $w(A_i)=h(\ell+1)+\ell=W-1$. We claim that every bundle $B$ that $i$ strictly prefers to $A_i$ has weight at least $W$. If $B$ contains at least $h+1$ high goods, these alone have weight at least $W$. Otherwise, let $H\le h$ be the number of high goods in $B$. Replacing values by weights reduces the value of each high good by $r-(\ell+1)>0$. Since $B$ has at most as many high goods as $A_i$, it loses at most as much as $A_i$ under this replacement. Hence $v_i(B)>v_i(A_i)$ implies $w(B)>w(A_i)=W-1$. The weights are integers, so $w(B)\ge W$.

It remains to bound the total weight available. A recipient of an $i$-high good in round $t$ has at most $h+1$ goods, each of weight at most $\ell+1$, so her bundle has weight at most $W$. For any other bundle $A_j$, write $H$ and $L$ for its numbers of high and low goods. Using $H+L\le h+\ell+1$ and $H\le h$, we obtain
\[
    w(A_j)=(H+L)+H\ell\le h+\ell+1+h\ell=W.
\]
Thus every assigned bundle has weight at most $W$, and $A_i$ has weight $W-1$. It follows that $w(M)\le nW-1$, whereas $n$ disjoint bundles strictly preferred to $A_i$ would require total weight at least $nW$. By \Cref{lem:bv-strict-packing}, $x\ge\mu_i$, proving condition~(b).

\paragraph{The envy guarantee.}
Under condition (a), every other bundle has value at most $x+1$. Removing any good reduces its value by at least $1$, so $i$ does not strongly envy it. In the remaining case $\ell+1<r$, a bundle that receives no $i$-high good in round $t$ still has value at most $x+1$. A bundle that receives such a good contains at most $h+1$ goods; removing any one leaves at most $h$ goods and hence value at most $hr\le x$. Thus, the envy guarantee also holds in condition (b).

\paragraph{No good is high for agent $i$.}
If $M=\varnothing$, both claims hold trivially. Otherwise, let $R$ be the final round.
In this case, $i$ is always unmatched, never marked, and active in every round. If she receives a good in the final round, she receives $R$ goods and $m\le nR$. If she misses a good in the final round, she receives $R-1$ goods and that final round assigns at most $n-1$ goods, so $m\le nR-1$. In either event she receives at least $\lfloor m/n\rfloor$ goods, which is her exact maximin share when each good has value $1$. Every other agent receives at most $R$ goods, proving the envy guarantee as well.
\end{proof}

\subsection{Auxiliary Weights and the Ordinal Guarantee}
\label{subsec:bv-weights}

\Cref{lem:bv-structure} proves that the complete allocation is \EFX{} and that the \MMS{} guarantee (thus any ($1$-out-of-$d$)-\MMS{} for $d\ge n$) is satisfied in condition (b). It remains to prove the ordinal guarantee in condition~(a). Fix an agent $i$ and write $x\coloneqq v_i(A_i)$.
We first show that an agent who receives no high good also receives her maximin share. For an agent who receives a high good, the argument depends on $x$. If $x\ge6$, the bound $v_i(A_j)\le x+1$ from \Cref{lem:bv-structure} gives $v_i(A_j)\le 7x/6$ for every $j\ne i$, and we can use the total value of the goods to prove the ordinal guarantee. If $x<6$, there are only finitely many possibilities for the numbers of high and low goods in $A_i$. For each possibility, we construct auxiliary weights that yield the same ordinal guarantee.

The following lemma formalizes the counting argument used in condition~(b) of \Cref{lem:bv-structure}. For a choice of non-negative item weights, write $w(B)\coloneqq\sum_{g\in B}w(g)$. We choose these weights so that every bundle that $i$ strictly prefers to $A_i$ has weight at least one. If the total weight of all goods is less than $d$, then every partition into $d$ bundles has a part of value at most $x$ to agent $i$.
The lemma shows how bounds on the weights of the assigned bundles give such a choice of $d$. The weights may depend on the agent and the case under consideration. All bundle comparisons and maximin shares are with respect to $v_i$.

\begin{lemma}
\label{lem:bv-weight-certificate}
Consider an agent $i$ and a complete allocation $A$. Write $x\coloneqq v_i(A_i)$. Assume there are non-negative item weights $w(\cdot)$ and a real number $c\ge1$ such that
\begin{enumerate}
    \item[(i)] $w(A_i)<1$;
    \item[(ii)] $w(A_j)\le c$ for every $j\ne i$; and
    \item[(iii)] $w(B)\ge1$ for every $B\subseteq M$ with $v_i(B)>x$.
\end{enumerate}
Then $x\ge\mu_i^d$ for $d=\lceil1+c(n-1)\rceil$.
\end{lemma}

\begin{proof}
Since the allocation is complete and from the first two conditions in the lemma statement, we have $w(M)=\sum_{j\in N}w(A_j)<1+c(n-1)\le d.$
Towards a contradiction, let us assume that $\mu_i^d>x$. Then there is a partition of $M$ into $d$ bundles, each of value strictly greater than $x$ to agent $i$. By condition~(iii), each bundle has weight at least $1$, so $w(M)\ge d$, contradicting $w(M)<d$. Thus $x\ge\mu_i^d$.
\end{proof}

The following example (\Cref{fig:weight-intuition}) illustrates the role of auxiliary weights.
Suppose $r=1.2$ and $A_i$ contains one high and one low good, so $x=2.2$. Give high goods weight $1/2$ and low goods weight $1/3$. Then $A_i$ has weight $5/6<1$, whereas every bundle worth more than $x$ contains either two high goods, one high and two low goods, or three low goods. Each of these combinations has weight at least $1$, as shown in \Cref{fig:weight-intuition}.
Thus, the weights give a common lower bound for all strictly better bundles, even though their values differ.

\begin{figure}[t]
    \centering
    \begin{tikzpicture}[
        x=1cm,y=1cm,
        font=\small,
        item/.style={draw=black,line width=.4pt},
    ]

    % Column headings
    \node[anchor=west] at (0,0.65) {Bundle};
    \node at (4.1,0.65) {Value};
    \node at (6.5,0.65) {Weight};

    % Own bundle: one high and one low good
    \filldraw[item,fill=black!12]
        (0,-0.25) rectangle ++(.65,.5);
    \node at (.325,0) {$1.2$};

    \draw[item] (.8,-0.25) rectangle ++(.65,.5);
    \node at (1.125,0) {$1$};

    \node[anchor=west] at (1.8,0) {$A_i$};
    \node at (4.1,0) {$x=2.2$};
    \node at (6.5,0) {$5/6<1$};

    % Two high goods
    \foreach \xx in {0,.8}{
        \filldraw[item,fill=black!12]
            (\xx,-1.45) rectangle ++(.65,.5);
        \node at (\xx+.325,-1.2) {$1.2$};
    }
    \node at (4.1,-1.2) {$2.4>x$};
    \node at (6.5,-1.2) {$1$};

    % One high and two low goods
    \filldraw[item,fill=black!12]
        (0,-2.45) rectangle ++(.65,.5);
    \node at (.325,-2.2) {$1.2$};

    \foreach \xx in {.8,1.6}{
        \draw[item] (\xx,-2.45) rectangle ++(.65,.5);
        \node at (\xx+.325,-2.2) {$1$};
    }
    \node at (4.1,-2.2) {$3.2>x$};
    \node at (6.5,-2.2) {$7/6$};

    % Three low goods
    \foreach \xx in {0,.8,1.6}{
        \draw[item] (\xx,-3.45) rectangle ++(.65,.5);
        \node at (\xx+.325,-3.2) {$1$};
    }
    \node at (4.1,-3.2) {$3>x$};
    \node at (6.5,-3.2) {$1$};

    \end{tikzpicture}
    \caption{Auxiliary weights when $r_i=1.2$, $w_H=1/2$, and $w_L=1/3$. Labels inside goods denote the values of the respective goods to agent $i$. Her own bundle (one high, one low) has weight below $1$.
    The remaining rows show all inclusion-wise minimal strictly better bundle types, each of weight at least $1$.}
    \label{fig:weight-intuition}
\end{figure}

When $A_i$ contains only low goods, we can choose weights so that the preceding argument applies with $d=n$, giving agent $i$ her exact \MMS{}.

\begin{lemma}
\label{lem:bv-low-only}
If the allocation procedure gives agent $i$ no high good, then $v_i(A_i)\ge\mu_i$.
\end{lemma}

The proof assigns suitable weights to high and low goods and applies the same counting argument with $d=n$. We give the details in \Cref{sec:appendix-bivalued}.

When $A_i$ contains a high good and has value below $6$, only finitely many pairs $(h,\ell)$ are possible. For each pair, we choose weights that separate $A_i$ from every strictly better bundle while keeping the weight of every other assigned bundle at most $\nicefrac{7}{6}$. The full table and verification are included in \Cref{app:bv-certificate}.

\begin{lemma}
\label{lem:bv-small-weights}
Consider $h\ge1$, $r>1$, and $x=hr+\ell<6$. If every other assigned bundle satisfies condition (a)of \Cref{lem:bv-structure}, then there are positive weights $w_H,w_L$ for high and low goods, respectively, such that
\[
    hw_H+\ell w_L<1,
    \qquad
    w(A_j)\le\frac76
    \quad(j\ne i),
\]
and $w(B)\ge1$ for every bundle $B$ with $v_i(B)>x$.
\end{lemma}

\begin{proof}
Since $r>1$ and $hr+\ell<6$, we have $h+\ell\le5$. Thus, there are fifteen possible pairs $(h,\ell)$ with $h\ge1$. For each pair, the relevant ratio range is $1<r<(6-\ell)/h$.

Every inclusion-wise minimal bundle of value more than $x$ contains at most six goods: any six goods already have value at least $6>x$. Every other assigned bundle also contains at most six goods, since its value is at most
$x+1<7$. It therefore suffices to consider the twenty-eight pairs
\[
    H,L\in\mathbb Z_{\ge0},
    \qquad H+L\le6.
\]

For each $(h,\ell)$, \Cref{tab:bv-weights} in \Cref{app:bv-certificate} divides the ratio range into intervals and specifies positive weights on each interval. The verification in \Cref{app:bv-certificate} shows that the own bundle has weight below $1$, every allowed assigned bundle has weight at most $7/6$, and every inclusion-wise minimal strictly better bundle has weight at least $1$.
Every strictly better bundle contains an inclusion-wise minimal strictly better bundle. Since all item weights are positive, adding the remaining goods cannot decrease its weight, so its weight is also at least $1$.
\end{proof}

We split the proof at $x=6$ because every other bundle is worth at most $x+1$ to agent $i$. When $x\ge6$, we have $x+1\le 7x/6$, so the total value of the goods gives the desired ordinal guarantee. When $x<6$, the auxiliary weights from the lemma give the same guarantee.
We will show the desired ordinal guarantee for $x\ge 6$ in the next section where we give the complete proof of \Cref{thm:bv-main}.

\subsection{Proof of \Cref{thm:bv-main}}

We first prove the result for positive valuations, then extend it to zero and constant values.

\subsubsection{Positive Valuations}

\begin{proof}[Proof of \Cref{thm:bv-main} for positive valuations]
We run the allocation procedure in \Cref{subsec:bv-algorithm} and set $d=\lceil(7n-1)/6\rceil$ (notice that $d\ge n$). The allocation is complete by \Cref{lem:bv-reachability} and \EFX{} by \Cref{lem:bv-structure}.

Consider an agent $i$, and use the notation $x=hr+\ell$ from above.
If condition (b) of \Cref{lem:bv-structure} applies, no good is high for $i$, or $h=0$, then \Cref{lem:bv-structure,lem:bv-low-only} give $x\ge\mu_i\ge\mu_i^d$. Otherwise, $h\ge1$ and condition (a) in \Cref{lem:bv-structure} holds. In particular, every other bundle is worth at most $x+1$ to $i$.

First, we consider the case where $x\ge6$. Completeness of the allocation implies
\[
\begin{aligned}
    v_i(M)
      &=\sum_{j\in N}v_i(A_j) \le x+(n-1)(x+1)=nx+n-1\le\left(n+\frac{n-1}{6}\right)x=\frac{7n-1}{6}x
      \le dx.
\end{aligned}
\]
There cannot be $d$ disjoint bundles each of value strictly greater than $x$, since their total value would exceed $dx$. Thus, by \Cref{lem:bv-strict-packing}, we have $x\ge\mu_i^d$.

If $x<6$, we apply \Cref{lem:bv-small-weights} and then \Cref{lem:bv-weight-certificate} with $c=7/6$. The resulting denominator is
\[
    \left\lceil1+\frac76(n-1)\right\rceil
       =\left\lceil\frac{7n-1}{6}\right\rceil
       =d.
\]
Hence $x\ge\mu_i^d$ in this case as well. Since the argument holds for every agent, the allocation is ($1$-out-of-$\left\lceil\frac{7n-1}{6}\right\rceil$)-\MMS{}. The polynomial running time follows from \Cref{lem:bv-reachability}.
\end{proof}

\subsubsection{Binary Values and Constant Valuations}
\label{subsec:bv-zero}

We now extend the construction to the non-negative valuations allowed in the preliminaries. The only additional cases are binary valuations with low value zero and valuations assigning the same value to every good. First, binary valuations have a useful property independent of our allocation procedure.

\begin{lemma}
\label{lem:bv-binary}
Consider agent $i$ with an additive valuation $v_i$ such that $v_i(g)\in\{0,b\}$ for every good $g$, where $b>0$. Then, the following statements hold.
\begin{enumerate}
    \item[(i)] If a complete allocation $A$ is \EFone{} for agent $i$ under $v_i$, then $v_i(A_i)\ge\mu_i$.

    \item[(ii)] Assume $m\ge1$. Define $\widetilde v_i$ by replacing each good value of $0$ with $b/(m+1)$ and keeping each good value of $b$ unchanged. If an allocation $A$ is \EFX{} for agent $i$ under $\widetilde v_i$, then the same allocation is \EFX{} for agent $i$ under $v_i$.
\end{enumerate}
\end{lemma}

For the first claim, \EFone{} ensures that the agent receives enough positive-valued goods to meet her maximin share.
For the second, the perturbations are small so any violation of \EFX{} under the original valuation remains a violation in the perturbed instance. The proof is deferred to \Cref{sec:appendix-bivalued}.

\begin{proof}[Completing the proof of \Cref{thm:bv-main}]
If $m=0$, give every agent the empty bundle. Otherwise, transform the valuations independently for each agent as follows. Keep each positive non-constant valuation unchanged. For an agent with values in $\{0,b_i\}$ and $b_i>0$, replace zero values by $b_i/(m+1)$; after normalization, her values lie in $\{1,m+1\}$. For a constant positive valuation, normalize every good value to $1$ and take $2$ as the high value not realized for any good. For an all-zero valuation, use the same all-ones valuation during the procedure.

On this profile, we run the positive-valuation procedure from \Cref{subsec:bv-algorithm}.
For unchanged positive valuations, the guarantees proved above apply directly; normalization does not affect them. For constant positive valuations, the no-high case of \Cref{lem:bv-structure} gives both \EFX{} and exact \MMS{}. For an all-zero valuation, any allocation is \EFX{} and \MMS{}.

For an originally binary valuation, \Cref{lem:bv-binary} transfers the \EFX{} comparisons from the transformed valuation to the original valuation. In particular, the complete allocation is also \EFone{} for that agent, so the same lemma gives her her exact maximin share. Since $d\ge n$, exact \MMS{} implies the required ordinal guarantee. These arguments apply separately to each agent, establishing both guarantees for the original profile. All transformations use rational numbers of polynomial bit length and preserve polynomial running time.
\end{proof}

\subsection{Tightness of Analysis}

The ordinal guarantee in \Cref{thm:bv-main} cannot be improved for \MatchAndPause{}, even when all agents have identical valuations. This does not rule out a stronger guarantee using a different algorithm.

\begin{proposition}
\label{prop:bv-tightness}
For every $n\ge2$, there is an instance with identical bivalued valuations on which \MatchAndPause{} returns an allocation that is not ($1$-out-of-$d'$)-\MMS{}, where
\[
    d'=\left\lceil\frac{7n-1}{6}\right\rceil-1.
\]
\end{proposition}

\begin{proof}
Consider $n$ agents with identical valuations. There are $n$ high goods, each of value $3/2$, and $2n-1$ low goods, each of value $1$.

In the first round, every agent is matched to a high good, so no agent is marked or paused. All high goods are allocated in this round. In the second round, every agent receives a low good. The remaining $n-1$ low goods are allocated in the third and final round. Thus, one agent $i$ receives one high good and one low good, giving her value $v_i(A_i)=5/2$. Every other agent receives one high good and two low goods.

We show that the goods can be partitioned into $d'$ bundles, each of value at least $3$. Since $d'=\left\lfloor\frac{7n-2}{6}\right\rfloor$, this will imply $\mu_i^{d'}\ge3>5/2=v_i(A_i)$.

If $n$ is even, group the high goods into pairs and form as many triples of low goods as possible.
Every such bundle has value $3$, and their number is
\[
    \frac n2+\left\lfloor\frac{2n-1}{3}\right\rfloor
    =\left\lfloor\frac{7n-2}{6}\right\rfloor=d'.
\]

If $n$ is odd, group all but one of the high goods into pairs. Combine the remaining high good with two low goods, and form as many triples as possible from the remaining $2n-3$ low goods. The pairs and triples have value $3$, and the mixed bundle has value $7/2$. The number of bundles is
\[
    \frac{n-1}{2}+1+
    \left\lfloor\frac{2n-3}{3}\right\rfloor
    =\left\lfloor\frac{7n-3}{6}\right\rfloor
    =\left\lfloor\frac{7n-2}{6}\right\rfloor=d',
\]
where the second equality uses that $n$ is odd.

In either case, we can add any leftover low goods to these bundles arbitrarily to obtain a partition of all goods.
Every bundle has a value at least $3$, proving that the output allocation is not ($1$-out-of-$d'$)-\MMS{}.
\end{proof}

\section{Conclusion and Future Directions}

We study allocations that simultaneously provide envy-based and ordinal maximin share guarantees. For general additive valuations, we establish the existence of complete \EFone{} allocations that are $1$-out-of-$\lfloor\varphi n\rfloor$-\MMS{}. For personalized bivalued valuations, we give a polynomial-time algorithm that achieves \EFX{} together with $1$-out-of-$\lceil(7n-1)/6\rceil$-\MMS{}.
Several questions remain. Can the ordinal guarantee be improved while preserving \EFone{} for general additive valuations?
Can our general additive guarantee be achieved in polynomial time?
For personalized bivalued valuations, a natural next question is whether a complete allocation can satisfy both \EFX{} and exact \MMS{}, or how close to exact \MMS{} one can get while ensuring \EFX{}.

\section*{AI Usage Disclosure}
The theoretical results were derived with assistance from GPT-6 Astra (max effort) over several rounds of interactions, wherein the authors provided intuition and guidance, narrowed the search for the approximation factor, and corrected intermediate errors. The authors manually verified the results, wrote the paper, and asked Claude Opus 5.5 and GPT-6 Astra to check and edit the arguments. The figures were generated by Claude Opus 5.5 from the authors' prompts and then modified for the paper. The authors take full responsibility for the content.

% \section*{Acknowledgments}
% This research was supported in part by NSF Awards IIS-2144413 and IIS-2107173. 
%We also thank the anonymous reviewers for their careful reading and constructive feedback, which improved the paper.

\bibliographystyle{plainnat}
\bibliography{References}

\clearpage

\appendix
\begin{center}
    \Large{\textbf{Appendix}}
\end{center}

\section{Missing Details from \Cref{sec:general-additive}} \label{sec:appendix-general-additive}

\begin{proof}[Proof of \Cref{lem:envy-free-matching}]
If $G$ admits a perfect matching, the claim follows immediately since there are no unmatched agents.

Otherwise, Hall's theorem gives a non-empty collection of bundles $\mathcal{Y}\subseteq\mathcal{B}$ with fewer neighboring agents than bundles, that is, $|\Gamma(\mathcal{Y})|<|\mathcal{Y}|$. 
Choose such a collection $\mathcal{Y}$ that contains no proper subset violating Hall's condition i.e. $|\Gamma(\mathcal{Z})|\ge|\mathcal{Z}|$ for every $\mathcal{Z}\subsetneq\mathcal{Y}.$ Such a collection exists because $\mathcal{B}$ is finite. Moreover, $|\mathcal{Y}|\ge 2$, since every individual bundle has at least one neighbor.

We first show that $\mathcal{Y}$ has exactly $|\mathcal{Y}|-1$ neighboring agents. Fix any bundle $B\in\mathcal{Y}$ and let $\mathcal{Y}'=\mathcal{Y}\setminus\{B\}$.
By minimality,
\[
    |\mathcal{Y}|-1
    =|\mathcal{Y}'|
    \le|\Gamma(\mathcal{Y}')|
    \le|\Gamma(\mathcal{Y})|
    <|\mathcal{Y}|.
\]
Since these cardinalities are integers, it follows that
\[
    |\Gamma(\mathcal{Y}')|
    =|\Gamma(\mathcal{Y})|
    =|\mathcal{Y}|-1.
\]

Now consider the bipartite graph between the bundles $\mathcal{Y}'$ and the agents $\Gamma(\mathcal{Y})$.
Every subset of $\mathcal{Y}'$ is a proper subset of $\mathcal{Y}$, so minimality of $\mathcal{Y}$ guarantees Hall's condition in this graph. Hence, there is a matching covering every bundle in $\mathcal{Y}'$. Since $|\mathcal{Y}'|=|\Gamma(\mathcal{Y})|$, this matching also covers every agent in $\Gamma(\mathcal{Y})$.

Use this matching in the original graph $G$, leaving all other vertices unmatched. Every unmatched agent lies outside $\Gamma(\mathcal{Y})$ and therefore has no edge to any bundle in $\mathcal{Y}$, including the matched bundles in $\mathcal{Y}'$.
Finally, the matching is non-empty because $|\mathcal{Y}'|=|\mathcal{Y}|-1\ge 1$.
\end{proof}

\begin{proof}[Proof of \Cref{lem:sparse-packing}]
We say that a good is \emph{large} if its value is at least $1$, it is \emph{medium} if its value lies in $(1/2,1)$, and is \emph{small} if its value lies in $(0,1/2]$. Without loss of generality, we can ignore zero-valued goods.
Like before, we assign the large goods as singleton bundles; if their number $h$ is at least $u$, we are done.
Otherwise, we initialize a bundle with a medium good and add small goods until its value reaches $1$ (the threshold), repeating the process till all medium goods are exhausted.
If the small goods run out before a bundle reaches $1$, then we complete that bundle with a second medium good, if available, and pair the remaining medium goods.
If all medium goods are used while small goods remain, we pack the remaining small goods greedily till the bundles reach the threshold.
We stop creating bundles as soon as $u$ acceptable bundles are formed.

Towards a contradiction, let us assume that the above procedure produces fewer than $u$ bundles. If small goods suffice to complete every bundle initialized with a medium good, then every completed bundle other than a large singleton has value below $3/2$ since, immediately before its last good is added, its value is below $1$, and the last good has value at most $1/2$. This also applies to bundles formed entirely from small goods.

Since fewer than $u$ bundles are produced, there are at most $u-h-1$ completed bundles other than the $h$ large singletons. Also, the final remaining goods together have a value below $1$.
Thus,
\[
    c_1(R)
    <h+\frac{3}{2}(u-h-1)+1
    \le\frac{3u-1}{2},
\]
contradicting~\eqref{eq:sparse-packing-condition}.
Otherwise, small goods run out during the phase where goods are added to the bundles initialized with a medium good. Let $a$ be the number of completed bundles containing one medium good, and let $b$ be the number containing two.
Their values are below $3/2$ and $2$, respectively, and there are no completed bundles with only small goods. Moreover, $h+a+b\le u-1$ (via our assumption for the contradiction).
If some goods remain unallocated, it must be one medium good and possibly small goods already attached to it, with total value below $1$. Thus, the number of medium-valued goods in $R$ is $r=a+2b+1$, and
\[
    c_1(R)
    <h+\frac32a+2b+1
    =(h+a+b)+\frac{r+1}{2}
    \le u+\frac{r-1}{2}.
\]
If no goods are left unassigned, then the number of medium-valued goods is $r=a+2b$, and
\[
    c_1(R)
    \le h+\frac32a+2b
    =(h+a+b)+\frac r2
    \le u-1+\frac r2
    <u+\frac{r-1}{2}.
\]
Both cases contradict~\eqref{eq:sparse-packing-condition}.
\end{proof}

\begin{proof}[Proof of \Cref{lem:anchor-packing}]

Consider an iteration and let the divider in this iteration be $i\in N'$.
Since agents with zero demand are already satisfied, $\tau_i>0$.
By~\eqref{eq:anchor-demands}, this implies that $i\in C$ and
\[
    0<\mu_i^d,
    \qquad
    \theta\mu_i^d\le v_i(f_i)<\mu_i^d.
\]

For this proof, normalize agent $i$'s valuation by her ordinal target, i.e. consider $v(R)\coloneqq\frac{v_i(R)}{\mu_i^d}$ for every $R\subseteq M$. Furthermore, define
\[
    a\coloneqq v(f_i),\qquad
    \delta\coloneqq1-a,\qquad
    k\coloneqq|S|,\qquad
    u\coloneqq n-k.
\]
Then $\theta\le a<1$, and the normalized demand is $\frac{\tau_i}{\mu_i^d}=\delta>0.$
It suffices to find $u$ pairwise disjoint bundles in $U$, each of normalized value at least $\delta$.
All capped values in this proof are taken with respect to $v$.

Since the normalized ordinal target is $1$, there is a partition of $M$ into $d$ bundles, each of value at least $1$. In particular, $c_1(M)\ge d.$

Let $b$ be the number of anchors worth more than $a$ to the divider.
Because $i$ is a sink of $H[N']$, all these anchors belong to agents in $S$.
Hence, $b\le k$.
Moreover, \Cref{lem:anchor-properties} implies that every non-anchor good has value at most $a$.
Finally, the invariant~\eqref{eq:unassigned-demand} gives
\[
    v(A_j)<\delta
    \qquad\text{for every }j\in S.
\]

We consider two cases based on the value of $a$. If $a\le 2/3$, then $a\le 2\delta$, so each bundle we form uses at most $2\delta$ value. If $a>2/3$, then $\delta < 1/3$. In this case, we start with a partition of $M$ into $d$ parts, each worth at least $1$, and obtain two bundles worth at least $\delta$ from each part that contains no anchor worth more than $a$.

\vspace{0.5em}

\underline{\emph{Case 1: $a\le2/3$.}}
We first bound the value remaining in $U$.
Each of the $b$ anchors worth more than $a$ contributes at most $1$ to capped value, and every other anchor contributes at most $a$. Thus, $c_1(F)\le b+(n-b)a$. The $k$ assigned supplementary bundles have total capped value at most $k\delta$.
Since the anchors, the assigned supplementary bundles, and $U$ partition $M$, we obtain
\[
    \begin{aligned}
    c_1(U)
    \ge d-b-(n-b)a-k\delta = d-na-(b+k)\delta \ge d-na-2k\delta,
    \end{aligned}
\]
where the last inequality uses $b\le k$.
Every good in $U$ is a non-anchor and has value at most $a<1$, so $v(U)=c_1(U)$.

Let $h$ be the number of goods in $U$ whose values are at least $\delta$.
If $h\ge u$, choose $u$ such goods as singleton bundles. Otherwise, reserve all $h$ of them as singletons, and let $R$ be the remaining goods in $U$.
Since each reserved singleton has value at most $a$, we have $v(R)\ge d-na-2k\delta-ha.$
Every good in $R$ has value below $\delta$.
Consequently, $c_\delta(R)=v(R)$.
To apply \Cref{lem:greedy-packing}, it suffices to show that $v(R)\ge\bigl(2(u-h)-1\bigr)\delta.$
Using $u=n-k$ and $\delta=1-a$, the difference between our lower bound ($d-na-2k\delta-ha$) and this required value ($(2(u-h)-1)\delta$) is
\[
    \begin{aligned}
    &d-na-2k\delta-ha
        -\bigl(2(u-h)-1\bigr)\delta\\
    &\qquad
        =d-na-ha-(2n-2h-1)(1-a)\\
    &\qquad
        =d-2n+1+(n-1)a+h(2-3a).
    \end{aligned}
\]
This expression is non-negative since $h(2-3a)\ge0$ (because $a\le2/3$) and
\[
    d-2n+1+(n-1)a
    \ge(n-1)(a-\theta)\ge0.
\]
The latter holds from the anchor conditions in the statement of the lemma ($d\ge 2n-1-(n-1)\theta$) and $a\ge\theta$.
Therefore, we can apply \Cref{lem:greedy-packing} to get $u-h$ bundles from $R$, each of value at least $\delta$.
Together with the $h$ reserved singletons, these give the required $u$ bundles.

\vspace{0.5em}

\underline{\emph{Case 2: $a>2/3$.}}
Here, $\delta=1-a<1/3$.
Consider a partition of $M$ into $d$ parts, each of value at least $1$ under $v$ (possible by definition of $\mu_i^d$).
We discard every part in this partition that contains an anchor of value more than $a$.
Since there are $b$ such anchors, at least $d-b$ parts remain.

Every good in a remaining part has value at most $a=1-\delta$ (for anchors, this follows from the discarded parts, where as for non-anchor goods, it follows from \Cref{lem:anchor-properties}).
We claim that each remaining part $P$ contains two disjoint bundles worth at least $\delta$.

If $P$ contains a good $g$ worth at least $\delta$, use $\{g\}$ as the first bundle.
Since $v(g)\le1-\delta$ and $v(P)\ge1$,
\[
    v(P\setminus\{g\})
    \ge1-(1-\delta)=\delta.
\]
Thus, the remainder of $P$ is a second acceptable bundle.

Otherwise, every good in $P$ has value below $\delta$.
Add goods greedily until the resulting bundle $B$ first reaches value $\delta$.
This is possible because $v(P)\ge1$, and $\delta\le v(B)<2\delta.$
The remainder satisfies
\[
    v(P\setminus B)>1-2\delta>\delta,
\]
where the last inequality uses $\delta<1/3$.
This again gives two disjoint acceptable bundles.

Applying this argument to the remaining parts produces at least $2(d-b)$ pairwise disjoint bundles, each worth at least $\delta$, none of which contains an anchor of value more than $a$.
Among the other $n-b$ anchors, each belongs to at most one of these bundles.
Discarding every bundle containing an anchor therefore leaves at least $2(d-b)-(n-b)=2d-n-b$ pairwise disjoint bundles consisting entirely of non-anchor goods.

Each surviving bundle has capped value at least $\delta$ at cap $\delta$.
Consequently,
\[
    c_\delta(M\setminus F)\ge(2d-n-b)\delta.
\]
We now remove the supplementary bundles already assigned to agents in $S$ to get capped value of $U$.
Each has value below $\delta$, and hence capped value below $\delta$.
It follows that
\[
    \begin{aligned}
    c_\delta(U)
    &\ge(2d-n-b-k)\delta\\
    &\ge(2d-n-2k)\delta\\
    &\ge\bigl(2(n-k)-1\bigr)\delta\\
    &=(2u-1)\delta.
    \end{aligned}
\]
The second inequality uses $b\le k$, and the third uses $2d\ge3n-1$.
By \Cref{lem:greedy-packing}, the pool $U$ contains $u$ pairwise disjoint bundles worth at least $\delta$.

Thus, in either case, the divider can form the requested bundles. Returning to the original valuation, each bundle has value at least $\delta\mu_i^d=\tau_i$ to agent $i$.
\end{proof}

\begin{proof}[Proof of \Cref{thm:golden-ratio-ef1}]
We first show the analysis for instances with $n\ge5$.
Let $\Delta\coloneqq\sqrt{5n^2-6n+5}$, $d\coloneqq D(n)$, and $K\coloneqq2n-d-1.$
Furthermore, choose $\theta$ and $q_0$ as in~\eqref{eq:chosen-parameters}.
We verify that these parameters satisfy all assumptions of \Cref{prop:ordinal-ef1-parameters}.

We begin by verifying the sparse-packing condition $2d\ge3n-1$ and the parameter restrictions $0<\theta\le1/2$ and $q_0\ge1$.
For our choices $\theta=K/(n-1)$ and $q_0=K-1$, it suffices to show
\[
    \frac{3n-1}{2}\le d\le2n-3,
\]
which also gives $2\le K\le(n-1)/2$.
By the definition of $d$, these bounds follow from $2n\le\Delta\le3n-5$. We establish these two inequalities by comparing squares. Since $n\ge5$,
\[
    \Delta^2-(2n)^2
    =n^2-6n+5
    =(n-1)(n-5)\ge0,
\]
and
\[
    (3n-5)^2-\Delta^2
    =4n^2-24n+20
    =4(n-1)(n-5)\ge0.
\]
Since the quantities being compared are non-negative, we have $2n\le\Delta\le3n-5$.
Using the definition of $d$, the lower bound on $\Delta$ gives
\[
    d
    \ge\frac{n-1+\Delta}{2}
    \ge\frac{3n-1}{2}.
\]
The upper bound gives
\[
    d
    \le
    \left\lceil\frac{n-1+(3n-5)}{2}\right\rceil
    =2n-3,
\] thereby proving the bounds on $K$ and \eqref{eq:parameter-sparse}. Thus, $d\ge n$, $0<\theta\le1/2$, and $q_0\ge1$. Also, $d$, $K$, and $q_0$ are integers. By our choices of $\theta$ and $q_0$,
\[
    2n-1-(n-1)\theta
    =2n-1-K=d,
\]
and
\[
    q_0+1=K=2n-d-1.
\]
Hence, \eqref{eq:parameter-anchor} and \eqref{eq:parameter-dividers} hold with equality. For the remaining condition, observe that
\[
    K
    =2n-1-
      \left\lceil\frac{n-1+\Delta}{2}\right\rceil
    =\left\lfloor\frac{3n-1-\Delta}{2}\right\rfloor
    =\lfloor\kappa(n)\rfloor.
\]

Therefore, $K\le\kappa(n)$, and \eqref{eq:parameter-quadratic} holds. For our chosen parameters, the difference between the left- and the right-hand side of \eqref{eq:parameter-automatic} can be written as
\[
    \begin{aligned}
    &d-n-(n-1)\theta-(1-\theta)(q_0-1)=
    \frac{K^2-(3n-1)K+n^2-1}{n-1},
    \end{aligned}
\]
 where the numerator is non-negative by \eqref{eq:parameter-quadratic}, and $n-1>0$; therefore, \eqref{eq:parameter-automatic} holds.

Since all assumptions of \Cref{prop:ordinal-ef1-parameters} are now satisfied, it follows that there exists a complete allocation that is both \EFone{} and ($1$-out-of-$D(n)$)-\MMS{}.

We next compare $D(n)$ with $\lfloor\varphi n\rfloor$. For $n\ge5$, $(\sqrt5\,n-1)^2-\Delta^2 =(6-2\sqrt5)n-4>0.$
Since both $\sqrt5\,n-1$ and $\Delta$ are positive, $\Delta<\sqrt5\,n-1$, whereby,
\[
    \frac{n-1+\Delta}{2}
    <
    \frac{(1+\sqrt5)n-2}{2}
    =\varphi n-1.
\]
Since $\lceil x\rceil<x+1$ for every real $x$,
\[
    D(n)<\varphi n.
\]
As $D(n)$ is an integer, this implies $D(n)\le\lfloor\varphi n\rfloor$.
The ordinal maximin share is non-increasing in its denominator, so an allocation where each agent satisfies $\mu_i^{D(n)}$ also satisfies $\mu_i^{\lfloor\varphi n\rfloor}$ for every agent $i$.

It remains to consider instances in which $n\le4$. For $n=1$, we have $D(1)=1=\lfloor \varphi n\rfloor$. Assigning all goods to the only agent gives both guarantees. For $n\in\{2,3,4\}$, choose $\theta=1/2$, $q_0=1$, and
\[
    d=
    \begin{cases}
        3, & n=2,\\
        4, & n=3,\\
        6, & n=4.
    \end{cases}
\]
For each of these values, we have $d=D(n)=\lfloor\varphi n\rfloor$. With $\theta=1/2$ and $q_0=1$, the first three conditions of \Cref{prop:ordinal-ef1-parameters} all reduce to
\[
    d\ge\frac{3n-1}{2},
\]
while the fourth reduces to $d\ge2n-3$. Both inequalities hold for each of the three choices, and $d\ge n$. The proposition therefore gives the required allocation for these values of $n$ as well.
\end{proof}

\begin{proof}[Proof of \Cref{lem:ef1-completion}]
Construct the envy graph with the agents as vertices and an edge $i\to j$ whenever $v_i(A_i)<v_i(A_j)$.
If this graph contains a directed cycle, reassign each bundle on the cycle to the agent who points to its owner. Every participating agent strictly improves in this reassignment.
Moreover, since the collection of assigned bundles is unchanged, and no agent's value for her own bundle decreases, every previous \EFone{} comparison remains valid for the reassigned bundles i.e. \EFone{} is preserved.

Once there is no directed cycle, choose an agent with indegree zero (source) and assign her an unassigned good.
No other agent previously envied her, so deleting the newly added good eliminates envy toward her new bundle i.e. every agent is \EFone{} towards her.
All other required comparisons remain valid.
Repeating this process assigns every good without decreasing any agent's value.

Finally, note that between assignments of new goods, cycle rotations can occur only finitely many times: every agent's own value is non-decreasing among the finitely many values of the current bundles, and each rotation strictly improves its participants.
\end{proof}

\section{Missing Details from \Cref{sec:bv}} \label{sec:appendix-bivalued}

\begin{proof}[Proof of \Cref{lem:bv-low-only}]
If no good in $M$ is high for $i$, then every good has value $1$ to her. As shown in the proof of \Cref{lem:bv-structure}, she receives at least $\lfloor m/n\rfloor$ goods, which is her maximin share.

Now assume that some good is high for $i$, but she receives none. Thus $h=0$ and $x=\ell$. Let $t$ be the last round in which a good high for $i$ is allocated, and let $k=\lfloor r\rfloor$. Since $i$ receives no high good in round $t$, the counting argument in Case~2 of the proof of \Cref{lem:bv-structure} applies. In particular, $h=t-1$, so $t=1$. Thus, all goods that are high for $i$ are allocated in a single round, and each bundle contains at most one such good.
By \eqref{eq:bv-round-count}, there are at most $\ell+1$ rounds, so a bundle containing only low goods has at most $\ell+1$ goods.
A bundle containing one high good has $H=1=h+1$, so \eqref{eq:bv-extra-high} bounds its number of low goods by $\max\{0,\ell-k\}$.

Assign weight $1$ to each low good and weight $w_H\coloneqq\min\{k+1,\ell+1\}$ to each high good. We claim that every assigned bundle has weight at most $\ell+1$. For a bundle containing only low goods, this follows from the bound on its size. For a bundle containing a high good, if $\ell<k$, it contains no low goods and has weight $\ell+1$. If $\ell\ge k$, its weight is at most $(k+1)+(\ell-k)=\ell+1$. Since $A_i$ itself has weight $\ell$, completeness of the allocation gives $w(M)\le n(\ell+1)-1$.

We now show that every bundle $B$ with $v_i(B)>x=\ell$ has weight at least $\ell+1$. A bundle containing only low goods must contain at least $\ell+1$ of them to have value greater than $\ell$. If $\ell<k$, then a single high good has weight $\ell+1$. 
If $\ell\ge k$, then each high good has weight $k+1\ge r$, and each low good has weight equal to its value. Thus $w(B)\ge v_i(B)>\ell$. Since $w(B)$ is an integer, it is at least $\ell+1$ in this case as well.

Towards a contradiction, assume $\mu_i>x$. Then there is a partition of $M$ into $n$ bundles, each of value strictly greater than $x$ to agent $i$. Each of these bundles has weight at least $\ell+1$, so $w(M)\ge n(\ell+1)$, contradicting $w(M)\le n(\ell+1)-1$. Hence $x\ge\mu_i$.
\end{proof}

\begin{proof}[Proof of \Cref{lem:bv-binary}]
Let $q$ be the number of goods the agent values at $b$, and let $h$ be the number of these goods in her own bundle. Her maximin share is $b\lfloor q/n\rfloor$: she can distribute the $q$ positive-valued goods as evenly as possible among $n$ parts, and no partition can give every part more than $\lfloor q/n\rfloor$ such goods.
The \EFone{} condition implies that every other bundle contains at most $h+1$ positive-valued goods. Consequently,
\[
    q\le h+(n-1)(h+1)
      =nh+n-1
      <n(h+1).
\]
Thus $\lfloor q/n\rfloor\le h$, proving the first claim.

For part~(ii), assume that $A$ is \EFX{} for agent $i$ under $\widetilde v_i$. If $A$ is not \EFX{} under $v_i$ for agent $i$, there would be some $j\ne i$ and some $g\in A_j$ for which $v_i(A_j\setminus\{g\})>v_i(A_i)$. Since both values are multiples of $b$, the gap is at least $b$.

Replacing each zero value by $b/(m+1)$ increases the value of $A_i$ by at most $mb/(m+1)<b$, since $A_i$ contains at most $m$ goods. The value of $A_j\setminus\{g\}$ can only increase. Thus,
\[
    \widetilde v_i(A_i)
    < v_i(A_i)+b
    \le v_i(A_j\setminus\{g\})
    \le \widetilde v_i(A_j\setminus\{g\}),
\]
contradicting \EFX{} under $\widetilde v_i$. Hence, $A$ is also \EFX{} for agent $i$ under $v_i$.
\end{proof}

\subsection{The Auxiliary-weight Construction}
\label{app:bv-certificate}

This appendix gives the finite verification used in \Cref{lem:bv-small-weights}. 
Consider a fixed pair $(h,\ell)$ with $h\ge1$ and $h+\ell\le5$. For each ratio $r$ in $(1,(6-\ell)/h)$, \Cref{tab:bv-weights} specifies positive weights $w_H$ and $w_L$. The fifth column is the own weight $hw_H+\ell w_L$. The last column bounds the weight of every possible other bundle.

\paragraph{The count pairs that must be checked.}
By the proof of \Cref{lem:bv-small-weights}, it suffices to examine the twenty-eight pairs
\[
    \mathcal P\coloneqq
       \{(H,L)\in\mathbb Z_{\ge0}^2:H+L\le6\}.
\]
To bound other-bundle weights, we consider the following two families of count pairs:
\begin{align}
    &H\le h,
    \qquad
    (H-h)r+L-\ell\le1;
    \label{eq:bv-normal-pattern}\\
    &H=h+1,
    \qquad
    L\le\max\{0,\ell-\lfloor r\rfloor\},
    \qquad
    r+L-\ell\le1.
    \label{eq:bv-extra-pattern}
\end{align}
Under condition~(a) of \Cref{lem:bv-structure}, every other bundle has either $H\le h$, in which case its count pair satisfies \eqref{eq:bv-normal-pattern}, or $H=h+1$, in which case its count pair satisfies \eqref{eq:bv-extra-pattern}.

\paragraph{Checking one row.}
Fix a row in the table with ratio interval $I$ and endpoints $\alpha\le\beta$. The following verifies every assertion needed for that row.
\begin{enumerate}
    \item[(i)]
    Substitute the listed weights to check $hw_H+\ell w_L<1$. Every entry in the fifth column is strictly below $1$.

    \item[(ii)]
    For each $(H,L)\in\mathcal P$ with $Hw_H+Lw_L<1$, check
    \[
        (H-h)\alpha+L-\ell\le0,
        \qquad
        (H-h)\beta+L-\ell\le0.
    \]
    The expression $(H-h)r+L-\ell$ is affine in $r$, so these endpoint inequalities imply $Hr+L\le hr+\ell$ throughout $I$. Checking both endpoints is valid even when an endpoint is excluded from $I$ since the inequalities presented here hold on the closure as well.

    \item[(iii)]
    Check all possible pairs in \eqref{eq:bv-normal-pattern}. If $H=h$, such a pair is possible precisely when $L\le\ell+1$. If $H<h$, it is possible for some $r\in I$ precisely when
    \[
        I\cap
        \left[\frac{L-\ell-1}{h-H},\infty\right)
        \ne\varnothing.
    \]
    For every possible pair, check that $Hw_H+Lw_L$ is at most the last column of the table.

    \item[(iv)]
    For \eqref{eq:bv-extra-pattern}, let $H=h+1$ and consider $k\in\{1,\ldots,5\}$ as a possible value of $\lfloor r\rfloor$.
    A pair is possible precisely when, for some such $k$,
    \[
        L\le\max\{0,\ell-k\},
        \qquad
        I\cap[k,k+1)\cap(-\infty,\ell+1-L]
        \ne\varnothing.
    \]
    For each possible pair, again check its weight against the last column. A singleton intersection is non-empty only when its point belongs to every intersecting interval.
\end{enumerate}

These tests involve only the listed rational numbers and the twenty-eight pairs in $\mathcal P$. They give the own weights and other-bundle bounds displayed in the table. The largest own weight is $29/30<1$, and the largest other-bundle weight is $7/6$. For each of the fifteen pairs $(h,\ell)$, the intervals cover $(1,(6-\ell)/h)$ without a gap, including every interior boundary point. Thus, the tests verify the weight construction for every admissible real ratio, not only for a finite sample of ratios.

\paragraph{An illustrative row.}
Consider $(h,\ell)=(1,1)$ and $1<r<2$. The table assigns $w_H=1/2$ and $w_L=1/3$, giving the own-bundle weight $5/6$.
A pair with weight below $1$ satisfies $3H+2L<6$. Its coordinate-wise maximal possibilities are $(0,2)$ and $(1,1)$, whose values are $2\le r+1$ and $r+1$, respectively. Hence, no strictly better bundle has weight below $1$.

For a pair in \eqref{eq:bv-normal-pattern}, if $H=0$, then $L\le r+2<4$, so $L\le3$ and its weight is at most $1$. If $H=1$, then $L\le2$ and its weight is at most $1/2+2/3=7/6$. An exceptional pair has $H=2$ and $L\le\max\{0,1-\lfloor r\rfloor\}=0$, so its weight is $1$. This verifies all three weight requirements on the entire interval.

\begingroup
\small
\renewcommand{\arraystretch}{1.42}
\setlength{\tabcolsep}{7pt}
\begin{longtable}{@{}cccccc@{}}
\caption{Auxiliary weights for $h\ge1$ and $hr+\ell<6$.
The last column is the maximum over the pairs described in
\eqref{eq:bv-normal-pattern} and \eqref{eq:bv-extra-pattern},
for some ratio in the displayed interval.}
\label{tab:bv-weights}\\
\toprule
$(h,\ell)$
& Ratio interval
& $w_H$
& $w_L$
& Own weight
& Other-bundle bound\\
\midrule
\endfirsthead

\multicolumn{6}{c}{\tablename\ \thetable\ (continued)}\\
\toprule
$(h,\ell)$
& Ratio interval
& $w_H$
& $w_L$
& Own weight
& Other-bundle bound\\
\midrule
\endhead

\midrule
\multicolumn{6}{r}{Continued on the next page}\\
\endfoot

\bottomrule
\endlastfoot

$(1,0)$
& $(1,2)$
& $\frac12$ & $\frac12$
& $\frac12$ & $1$\\

$(1,0)$
& $[2,3)$
& $\frac23$ & $\frac13$
& $\frac23$ & $1$\\

$(1,0)$
& $[3,4)$
& $\frac34$ & $\frac14$
& $\frac34$ & $1$\\

$(1,0)$
& $[4,5)$
& $\frac45$ & $\frac15$
& $\frac45$ & $1$\\

$(1,0)$
& $[5,6)$
& $\frac56$ & $\frac16$
& $\frac56$ & $1$\\

$(1,1)$
& $(1,2)$
& $\frac12$ & $\frac13$
& $\frac56$ & $\frac76$\\

$(1,1)$
& $[2,3)$
& $\frac12$ & $\frac14$
& $\frac34$ & $1$\\

$(1,1)$
& $[3,4)$
& $\frac35$ & $\frac15$
& $\frac45$ & $1$\\

$(1,1)$
& $[4,5)$
& $\frac23$ & $\frac16$
& $\frac56$ & $1$\\

$(1,2)$
& $(1,2)$
& $\frac38$ & $\frac14$
& $\frac78$ & $\frac98$\\

$(1,2)$
& $\{2\}$
& $\frac25$ & $\frac15$
& $\frac45$ & $1$\\

$(1,2)$
& $(2,3)$
& $\frac12$ & $\frac15$
& $\frac9{10}$ & $\frac{11}{10}$\\

$(1,2)$
& $[3,4)$
& $\frac12$ & $\frac16$
& $\frac56$ & $1$\\

$(1,3)$
& $(1,\frac32]$
& $\frac3{10}$ & $\frac15$
& $\frac9{10}$ & $\frac{11}{10}$\\

$(1,3)$
& $(\frac32,2)$
& $\frac13$ & $\frac15$
& $\frac{14}{15}$ & $\frac{17}{15}$\\

$(1,3)$
& $\{2\}$
& $\frac13$ & $\frac16$
& $\frac56$ & $1$\\

$(1,3)$
& $(2,3)$
& $\frac5{12}$ & $\frac16$
& $\frac{11}{12}$ & $\frac{13}{12}$\\

$(1,4)$
& $(1,\frac32]$
& $\frac14$ & $\frac16$
& $\frac{11}{12}$ & $\frac{13}{12}$\\

$(1,4)$
& $(\frac32,2)$
& $\frac5{18}$ & $\frac16$
& $\frac{17}{18}$ & $\frac{10}{9}$\\

$(2,0)$
& $(1,\frac32)$
& $\frac13$ & $\frac13$
& $\frac23$ & $1$\\

$(2,0)$
& $[\frac32,2)$
& $\frac37$ & $\frac27$
& $\frac67$ & $\frac87$\\

$(2,0)$
& $[2,\frac52)$
& $\frac25$ & $\frac15$
& $\frac45$ & $1$\\

$(2,0)$
& $[\frac52,3)$
& $\frac5{11}$ & $\frac2{11}$
& $\frac{10}{11}$ & $\frac{12}{11}$\\

$(2,1)$
& $(1,\frac32)$
& $\frac13$ & $\frac14$
& $\frac{11}{12}$ & $\frac76$\\

$(2,1)$
& $[\frac32,2)$
& $\frac13$ & $\frac29$
& $\frac89$ & $\frac{10}{9}$\\

$(2,1)$
& $[2,\frac52)$
& $\frac13$ & $\frac16$
& $\frac56$ & $1$\\

$(2,2)$
& $(1,\frac32)$
& $\frac4{15}$ & $\frac15$
& $\frac{14}{15}$ & $\frac{17}{15}$\\

$(2,2)$
& $[\frac32,2)$
& $\frac3{11}$ & $\frac2{11}$
& $\frac{10}{11}$ & $\frac{12}{11}$\\

$(2,3)$
& $(1,\frac32)$
& $\frac29$ & $\frac16$
& $\frac{17}{18}$ & $\frac{10}{9}$\\

$(3,0)$
& $(1,\frac43)$
& $\frac14$ & $\frac14$
& $\frac34$ & $1$\\

$(3,0)$
& $[\frac43,\frac32)$
& $\frac4{13}$ & $\frac3{13}$
& $\frac{12}{13}$ & $\frac{15}{13}$\\

$(3,0)$
& $[\frac32,\frac53)$
& $\frac3{10}$ & $\frac15$
& $\frac9{10}$ & $\frac{11}{10}$\\

$(3,0)$
& $[\frac53,2)$
& $\frac5{16}$ & $\frac3{16}$
& $\frac{15}{16}$ & $\frac98$\\

$(3,1)$
& $(1,\frac43)$
& $\frac14$ & $\frac15$
& $\frac{19}{20}$ & $\frac{23}{20}$\\

$(3,1)$
& $[\frac43,\frac32)$
& $\frac14$ & $\frac3{16}$
& $\frac{15}{16}$ & $\frac98$\\

$(3,1)$
& $[\frac32,\frac53)$
& $\frac14$ & $\frac16$
& $\frac{11}{12}$ & $\frac{13}{12}$\\

$(3,2)$
& $(1,\frac43)$
& $\frac5{24}$ & $\frac16$
& $\frac{23}{24}$ & $\frac98$\\

$(4,0)$
& $(1,\frac54)$
& $\frac15$ & $\frac15$
& $\frac45$ & $1$\\

$(4,0)$
& $[\frac54,\frac43)$
& $\frac5{21}$ & $\frac4{21}$
& $\frac{20}{21}$ & $\frac87$\\

$(4,0)$
& $[\frac43,\frac32)$
& $\frac4{17}$ & $\frac3{17}$
& $\frac{16}{17}$ & $\frac{19}{17}$\\

$(4,1)$
& $(1,\frac54)$
& $\frac15$ & $\frac16$
& $\frac{29}{30}$ & $\frac{17}{15}$\\

$(5,0)$
& $(1,\frac65)$
& $\frac16$ & $\frac16$
& $\frac56$ & $1$\\

\end{longtable}
\endgroup

\end{document}